\documentclass[twocolumn]{aastex62}

\usepackage{amsmath, mathtools, amsthm, amssymb}
\usepackage{enumitem}
\usepackage{footmisc}
\hypersetup{colorlinks=true, allcolors=cyan}
\received{XX}
\revised{YY}
\accepted{ZZ}
\submitjournal{ApJ}

\shorttitle{Performance Enhancement of FoF Galaxy-finder}
\shortauthors{Rhee et al.}

\begin{document}
\title{Performance Enhancement of Tree-based Friends-of-friend Galaxy-finder for High-resolution Simulations of Galaxy Formation}

\email{jinsu.rhee@yonsei.ac.kr}

\author{Jinsu Rhee}
\affil{Department of Astronomy and Yonsei University Observatory, Yonsei University, Seoul 03722, Korea}

\author{Pascal Elahi}
\affil{Pawsey Supercomputing Research Centre, 1 Bryce Ave, Kensington WA 6151, Australia}

\author{Sukyoung K. Yi}
\affil{Department of Astronomy and Yonsei University Observatory, Yonsei University, Seoul 03722, Korea}


\def\NH{\textsc{NewHorizon}}
\def\VR{\textsc{VELOCIraptor}}
\def\BF{${\rm B}$}
\newcommand{\js}{\textcolor{blue}}


\begin{abstract}
Cosmological simulations are useful tools for studying the evolution of galaxies, and it is critical to accurately identify galaxies and their halos from raw simulation data.
The friends-of-friend (FoF) algorithm has been widely adopted for this purpose because of its simplicity and expandability to higher dimensions.
However, it is cost-inefficient when applied to high-resolution simulations because standard FoF implementation leads to too many distance calculations in dense regions.
We confirm this through our exercise of applying the 6-dimensional (6D) FoF galaxy finder code, \VR\ \citep[][]{Elahi19a}, on the \NH\ simulation \citep[][]{Dubois21}.
The high particle resolution of \NH\ ($M_{\rm star} \sim 10^4 M_{\odot}$) allows a large central number density ($10^{6}\,{\rm kpc}^{-3}$) for typical galaxies, resulting in a few days to weeks of galaxy searches for just one snapshot.
Even worse, we observed a significant decrease in the FoF performance in the high-dimensional 6D searches: ``the curse of dimensionality'' problem.
To overcome these issues, we have developed several implementations that can be readily applied to any tree-based FoF code.
They include limiting visits to tree nodes, re-ordering the list of particles for searching neighbor particles, and altering the tree structure.
Compared to the run with the original code, the new run with these implementations results in the identical galaxy detection with the ideal performance, $O(N \log{N})$, $N$ being the number of particles in a galaxy---with a speed gain of a factor of 2700 in 3D or 12 in 6D FoF search.
\end{abstract}



\section[]{Introduction}
\label{sec:Int}
Halos and galaxies are the fundamental building blocks of the Universe.
Their dynamics in groups and clusters, the kinematics of stars, gas, and dark matter particles inside them, and their integrated properties have been important subjects in the study of galaxy evolution.
While studies of halos and galaxies have mostly relied on observational data in the past, numerical simulations these days have successfully reproduced some of the key properties of galaxies and halos from the primordial density fluctuation \citep[e.g.,][]{Springel05a, Dubois14, Vogelsberger14, Schaye15, Hopkins18}, allowing us to study them within the computational laboratory.

Just as it is difficult and ambiguous to define and measure halos and galaxies in observations, there has been no consensus on how to identify them most accurately from raw simulation data \citep[e.g.,][]{Knebe11, Knebe13a, Knebe13b}.
This is because halos and galaxies are represented as a clustering of dark matter and/or stellar particles in a simulation, and their definition relies on the clustering algorithm used.
Even with a fixed algorithm, the free parameters within the algorithm often result in different systematic statistics for halos/galaxies \citep[e.g.,][]{More11}.

Most of the halo and galaxy finding codes adopt either the spherical over-density (SO) method \citep[][]{PS74} or the friends-of-friend (FoF) method \citep[][]{Davis85}. 
The SO method is primarily based on the theoretical prediction that a virialized halo has a characteristic mean density.
For example, in the Einstein-de Sitter Universe, the mean density of a halo is $18\pi^2\,\rho_{\rm mean}$, where $\rho_{\rm mean}$ is the mean density of the Universe.
Accordingly, the codes adopting the SO method (\citealp[SUBFIND;][]{Springel01,Dolag09}, \citealp[AdaptaHOP;][]{Aubert04}, \citealp[AHF;][just for the recent reference]{KK09}) first compute the density field of dark matter (stellar) particles, locate the peak density, and then identify halos (galaxies) as a spherical or ellipsoidal region centered at the peak density where the density criterion is satisfied.

On the other hand, FoF codes \citep[e.g.,][]{Behroozi13, Elahi19a} are purely based on the proximity among particles: two particles are linked if their metric (usually Euclidean distance) is less than a predefined ``linking length''.
An FoF group then corresponds to a set of particles linked to each other.
In this sense, FoF groups are free from the shape assumptions and hence can be considered more versatile, and in principle they do not have to depend on the cosmological definition of virialized objects.
In addition, FoF is easily expanded to more information than just configuration-space hence has the advantage of expandability.

However, from a computational perspective, FoF algorithms are more expensive than typical SO methods.
SO-based codes utilize a method of computing the density field via kernel smoothing \citep[e.g.,][]{Stadel01,Springel01,Aubert04} or a spatial grid \citep[e.g.,][]{GB94,KK09,Creasey18}, which ensures efficient calculations, whereas the FoF algorithm requires the metric calculation of all neighboring particles (neighbor query).
Therefore, a large number of metric calculations are required in over-dense regions, resulting in a poor execution speed.

Even worse, cosmological $N$-body simulations have progressively higher particle and spatial resolutions.
Therefore, dense regions are naturally constituted by a larger number of particles than before, thereby harming the efficiency of FoF-based codes.
For example, the \NH\ simulation \citep[]{Dubois21} used in this study is a cosmological zoom-in hydrodynamic simulation (see Section \ref{sec:Sample-Simulations}) with a high stellar particle resolution ($\sim 10^{4}M_{\odot}$), and its massive galaxies are often composed of $\gtrsim 10^{7}$ particles.
This sets the stage for the test performance of the FoF algorithm (Section \ref{sec:Background}).

Some recent FoF-based codes \citep[e.g.,][]{Kwon10,Behroozi13,FM17,Elahi19a} employ the $k$-dimensional tree (KD tree) structure \citep[][]{Bentley75, Moore01} when computing the neighbor queries, which significantly reduces the number of calculations required during neighbor queries (see Section \ref{sec:Background} for a detailed description).
Despite the use of the KD tree structure, a large number of metric calculations in over-dense regions are still required \citep[e.g.,][]{Behroozi13, FM17} in simulations with high particle and spatial resolutions (see Section \ref{sec:Background} for details).

Many recent public FoF codes have attempted to address the issue of low efficiency.
Mostly used is a parallelization manner (\citealp[ROCKSTAR;][]{Behroozi13}, \citealp[pFoF;][]{Roy14}, \citealp[\VR ;][just for the recent reference]{Elahi19a}): the entire volume is divided into subdomains, and each CPU performs a local FoF search in each subdomain.
As a result, the volume-wise parallelization shows a high level of scalability, alleviating the low-efficiency problem \citep[e.g.,][]{Behroozi13}.

However, several issues are associated with the poor efficiency of the volume-wise parallelization.
For example, an FoF group across multiple domains can be identified as individual groups in each domain.
After a local search, the work of stitching each FoF group into a larger one is required, which sometimes results in time-consuming calculations.
A subdomain with a massive FoF group is usually a bottleneck for calculations, and this is also a major cause of the poor efficiency.
In addition, searching for FoF groups in higher dimensions (e.g., 6-dimensional phase-space) often leads to a degradation of efficiency because of an issue known as ``the curse of dimensionality'' (see Section \ref{sec:Results-6DFoF} for details).

Many codes have attempted to perform FoF with better efficiency in various ways, such as using modified tree structures \citep[][]{Kwon10, FM17}, searching FoF groups through sub-sampling \citep[][]{Behroozi13}, or employing spatial grids \citep[][]{Creasey18}.
However, finding unique (without sub-sampling) solutions for searching FoF groups with practical performance on a growingly higher-resolution simulation is still a significant challenge.
Therefore, the purpose of this study is to find and provide implementations for finding ``exact'' FoF groups with good efficiency.
We highlight that our implementations show impressive performance gains with no or little trade-off other computing resources (e.g., memory) and can be easily applied to any FoF code based on the KD tree through a minor code modification.

As the main tool for improving the FoF efficiency, we utilize a phase-space halo/galaxy finder code, \VR \citep[][]{Elahi19a}.
The advantage of using \VR\ for the performance test is that the code finds exact 3-dimensional (3D) and 6-dimensional (6D) FoF groups without approximations.
The code is also able to read various input types (\citealp[\textsc{Ramses};][]{Teyssier02}, \citealp[\textsc{Gadget};][HDF5 data format, etc.]{Springel05b}) and multiple particle types with different masses and thus has a high level of expandability.
For example, the code has been successfully used in many studies \citep[e.g.,][just for the recent reference]{Elahi18, Canas19, Borrow20, Power20, Wright20, Bakels21}.

Our paper is organized as follows.
In Section \ref{sec:Sample}, we provide brief descriptions of the halo/galaxy finding code we employed (Section \ref{sec:Sample-VR}) and a simulation data sample (Section \ref{sec:Sample-Simulations}).
In Section \ref{sec:Background}, we address the problem of the performance degradation of FoF that occurs in a high-resolution simulation (Section \ref{sec:Background}).
In Section \ref{sec:Imple}, we describe the main implementations to overcome it.
In Section \ref{sec:Results}, we show the performance improvement of 3D (Section \ref{sec:Results-3DFoF}) and 6D FoF searches (Section \ref{sec:Results-6DFoF}) and performance results based on different tree structures (Section \ref{sec:Results-Tree}).
In Section \ref{sec:Conclusion}, we present our conclusions and provide a combination of implementations that shows the best performance.
In Appendix \ref{sec:Appendix}, we present the results for a different epoch (snapshot of the simulation) for reference.
We used a 48-core Intel CPU cluster\footnote{Four sets of Intel Xeon Processor E7-8857 v2 Family} running \VR\ with one MPI task and six OpenMP threads.
Parallel computations are performed in a way that each thread searches FoF groups in a specific local volume, i.e., we note that each FoF group is identified using a single thread alone.


\section[]{Sample}
\label{sec:Sample}

\subsection{VELOCIRaptor-STF}
\label{sec:Sample-VR}

We use \VR\footnote{\dataset[https://github.com/pelahi/VELOCIraptor-STF]{https://github.com/pelahi/VELOCIraptor-STF}\label{VRgithub}} \citep[][]{Elahi19a} as the main tool for identifying galaxies.
\VR\ finds galaxy entirely based on the FoF algorithm.
The code first identifies 3D FoF groups as a candidate for galaxies.
The particles of each 3D FoF group are mutually linked under the condition,
\begin{equation}
	\frac{|\vec{X_{\rm i}} - \vec{X_{\rm j}}|^2}{(b_{\rm 3D}\, \bar{d})^2} <1,
\end{equation}
where $\vec{X_{\rm i}}$ and $\vec{X_{\rm j}}$ are the positional vectors of the two particles, and $\bar{d}$ is the mean inter-particle distance.
The free parameter $b_{\rm 3D}$ is chosen to be 0.2, for which the FoF group corresponds to a typical virialized dark matter halo.

Subsequently, a 6D FoF search is performed to identify the substructures within each 3D FoF group.
In \VR , the 6D metric of the two particles $d_{\rm i, j}$ is defined as
\begin{equation}
	d_{\rm i, j} =\sqrt{ \frac{|\vec{X_{\rm i}} - \vec{X_{\rm j}}|^2}{(b_{\rm 6D}\, b_{\rm 3D}\, \bar{d})^2} + \frac{|\vec{V_{\rm i}} - \vec{V_{\rm j}} |^2}{(\alpha_{\rm 6D} \sigma_{\rm v})^2} },
\end{equation}
where $\vec{V_{\rm i}}$ and $\vec{V_{\rm j}}$ are the velocity vectors.
The new factors ($b_{\rm 6D}$, $\alpha_{\rm 6D}$, and $\sigma_{\rm v}$) are introduced above to properly normalize the 6D metric.
An additional normalization parameter for the configuration-space ($b_{\rm 6D}$) is employed and set as 0.2 in this study, motivated by the fact that galaxies in each 3D FoF group are more compact objects, therefore, a smaller physical linking length should be adopted \citep[][]{Canas19}.
The normalization of the velocity-space consists of two factors: the velocity dispersion of the parent 3D halo ($\sigma_{\rm v}$) and the free parameter ($\alpha_{\rm 6D}$).
Here, $\alpha_{\rm 6D}$ is empirically defined as unity: two particles at the same position are not linked if their relative velocity is greater than the dispersion of their system.
Then, particles with $d_{\rm i, j} < 1$ are tagged to the same 6D FoF group.

\VR\ also provides more sophisticated processing to identify substructures within a 6D FoF group that may be in the stage of merging or orbiting satellites.
Several FoF codes use their own method for substructure identification \citep[e.g.,][]{Behroozi13}.
Substructure identification in \VR\ is conducted by catching dynamically cold cores within each 6D FoF group iteratively using smaller linking lengths \citep[][]{Canas19, Elahi19a}.
As our study is focused on the performance of the FoF algorithm using a KD tree, we do not explore the performance of the substructure search.
Thus, the overall analysis is limited to the efficiency of the 3D FoF search and the 6D FoF search following the 3D FoF search.

\VR\ uses a KD tree algorithm \citep[][]{Bentley75} when finding FoF groups.
The KD tree structure has been widely used in many FoF codes and functions to significantly reduce the number of computations required to carry out a neighbor query.
In the original version of \VR , the KD tree was built in a balanced manner.
The two child nodes have the same number of particles, and the splitting dimension is defined as the dimension along which the parent node has the maximum length.

\subsection{Simulations}
\label{sec:Sample-Simulations}

The \NH\ data \citep[][]{Dubois21} are used in this study as a simulation sample.
The \NH\ simulation is a state-of-the-art cosmological zoom-in hydrodynamic simulation with high spatial and particle resolutions.
Herein, we provide a brief summary.
\NH\ is a zoom-in simulation from the parent simulation, {\textsc{Horizon-AGN}} \citep[][]{Dubois14}, a cosmological hydrodynamic simulation performed in a box of $(142\,\rm{cMpc})^3$.
Both simulations are run using the adaptive mesh refinement code, {\textsc{Ramses}} \citep[][]{Teyssier02}.
A ``field'' environment of a sphere with a radius of $10\,\rm{cMpc}$ is chosen from the {\textsc{Horizon-AGN}} volume as the domain of \NH , and that domain has been re-simulated with higher spatial and particle resolutions.
The dark matter particles with the highest resolution have a mass of $1.2 \times 10^{6}\,M_{\odot}$, and the stellar particle mass is approximately $10^{4}\,M_{\odot}$.
Thus, massive galaxies often have more than $10^{7}$ stellar particles.
In the \NH\ volume, the best spatial resolution reaches $34\,\rm{pc}$ at $z = 0$.
The simulation uses the WMAP-7 \citep[][]{Komatsu11} cosmology: $\Omega_{\rm m} = 0.272$, $\Omega_{\rm \lambda} = 0.728$, $\sigma_{\rm 8} = 0.81$,  $n_{\rm s} = 0.967$, and $H_{\rm 0} = 70.4\,\rm{km\,s^{-1}}\rm{Mpc^{-1}}$.

Throughout the analysis, we make use of a snapshot with a scale factor of $a = 0.289$ ($z = 2.46$) to find galaxies.
The choice of snapshot addresses the competing directions of having galaxies composed of many particles and avoiding excessive computation time needed by the original \VR\ at late cosmological times.
We also give the overall performance results at $a = 0.462$ in Appendix \ref{sec:Appendix}.
	

\section[]{Background}
\label{sec:Background}

\begin{figure}
\centering
\includegraphics[width=0.45\textwidth]{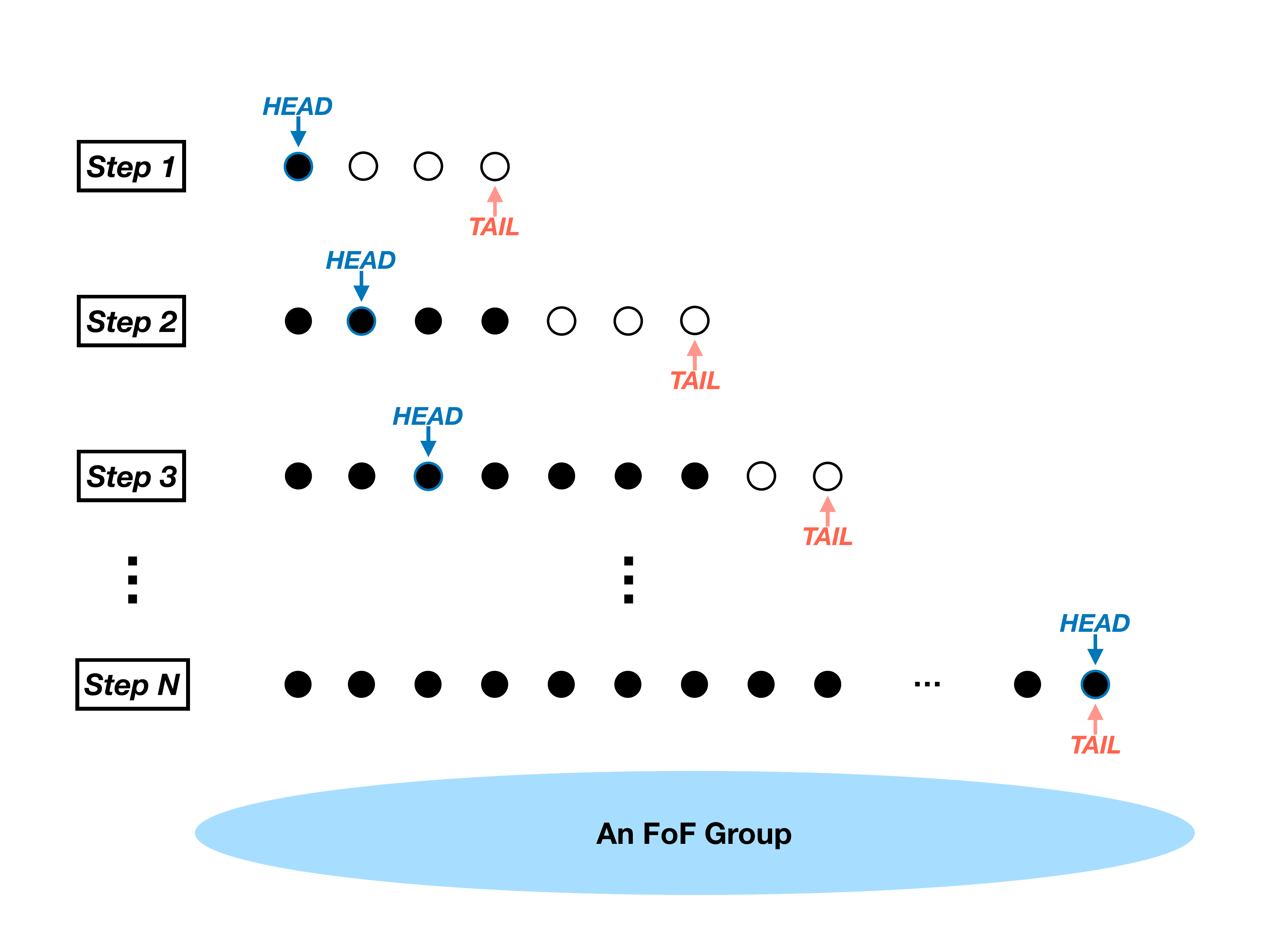}
\caption{Schematic illustration of the linked-list representation method.
Circles represent particles of the FoF group in each step, wherein open circles are the newly added particles in the step.
In each step, neighbor particles of the head particle are searched, and the tail particle is defined as the last particle in the list.
A particle next to the previous head particle becomes the new next head particle in the following step.
Searching for the exact FoF group is finished when the head and tail particles become identical.}
\label{fig:fig LLR}
\end{figure}

\begin{figure}
\includegraphics[width=0.45\textwidth]{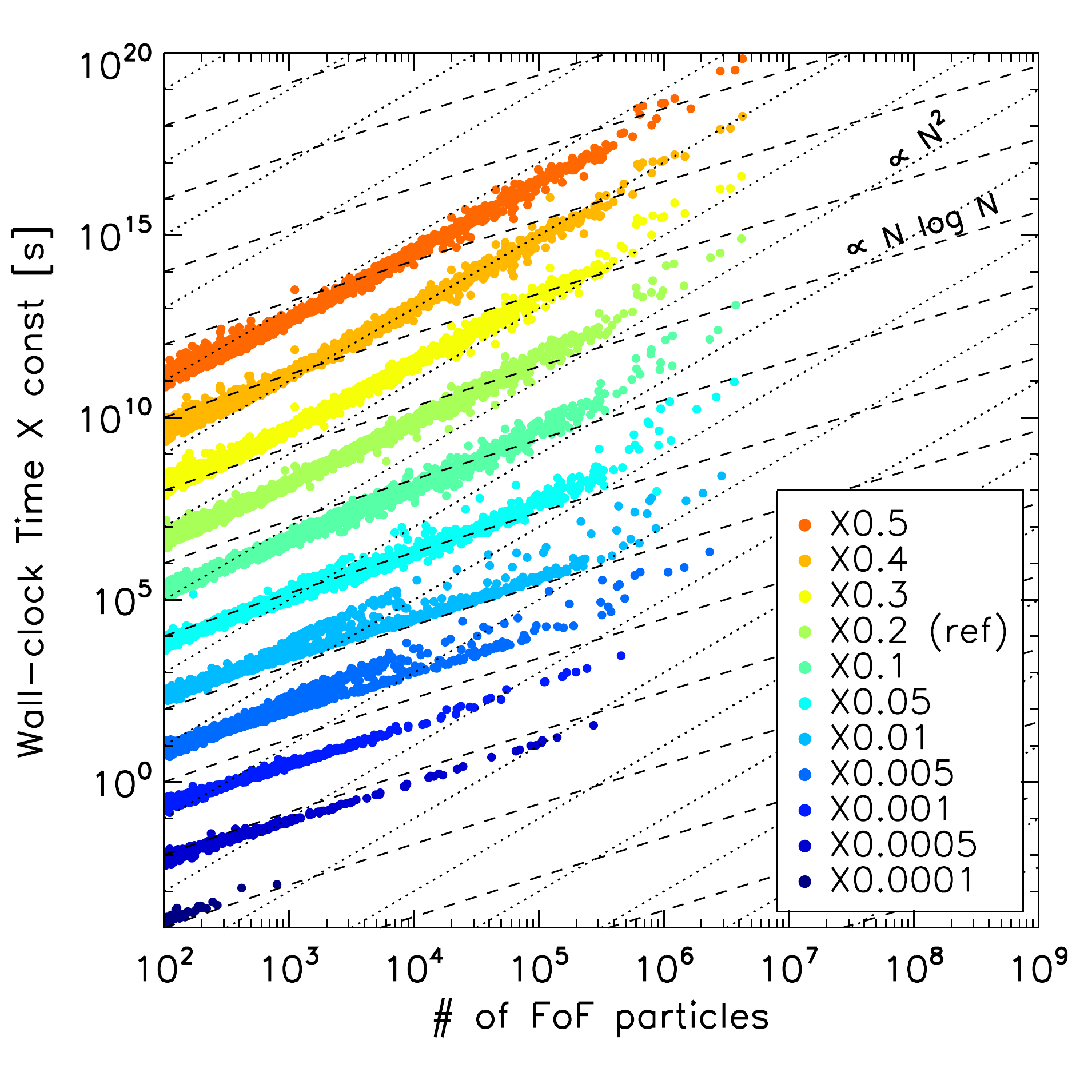}
\caption{
Wall clock-times of 3D FoF searches with different values of linking length.
Different linking lengths are used from 0.0001 to 0.5 times to the mean inter-particle distance of dark matter particles.
To separate each result for the illustrative purpose, we multiply a constant to each wall-clock time result, which results in the vertical offset in the plot.
The background dashed and dotted lines indicate the ideal ($\propto N \log N$) and the poor ($\propto N^2$) scaling relations, respectively.
Results with smaller linking length ($\times 0.0001 - \times 0.0005$) show the ideal scaling ($\propto N \log N$) whereas some of 3D FoF groups in the runs with longer linking length ($\times 0.001 - \times 0.5$) follow the poor scaling ($\propto N^2$).}
\label{fig:fig ltest}
\end{figure}

\begin{figure*}
\centering
\includegraphics[width=0.95\textwidth]{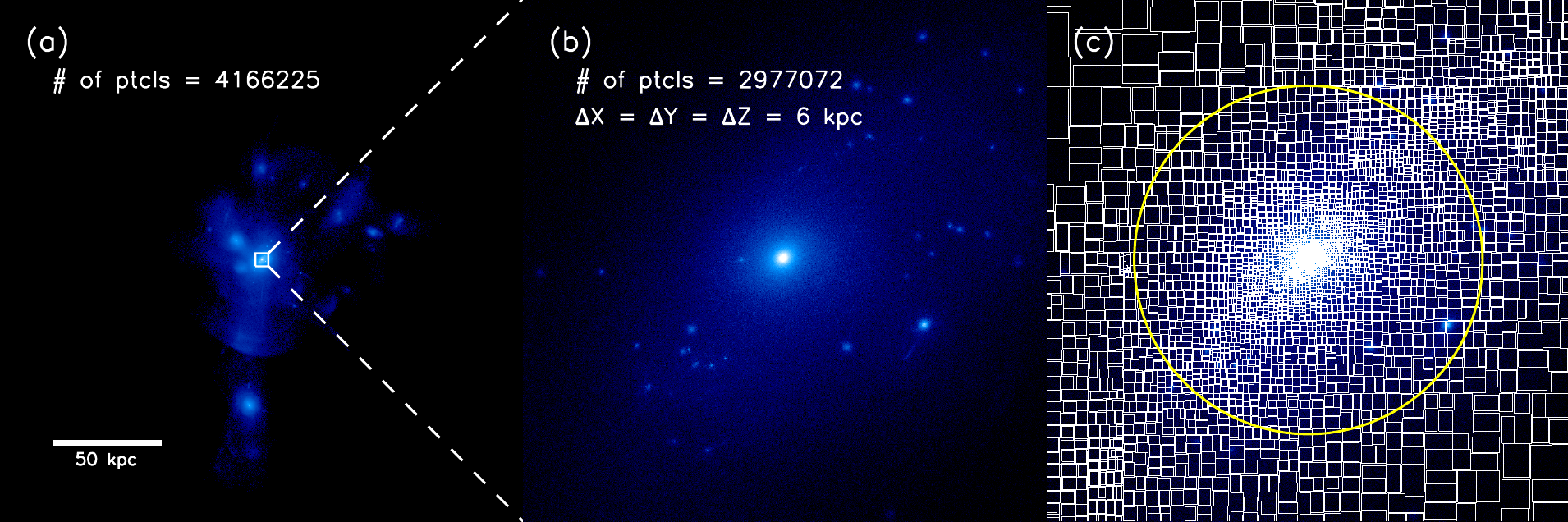}
\caption{Projected mass density of the most massive 3D FoF group at $a = 0.289$.
In Panel (a), the most massive 3D FoF group is shown, and its central region (white square; $6\,\rm{kpc}$ in length) is enlarged in Panel (b).
Most of the particles ($> 70\,\%$) are located in the central region.
In Panel (c), the leaf nodes of the KD tree are over-plotted to the central region, showing that the central area is composed of many tree nodes.
A search sphere of the radius of the linking length ($1.998\,\rm{kpc}$) is shown as an orange circle.}
\label{fig:fig GalandNode}
\end{figure*}

	FoF clustering corresponds to a set of maximum connected components (MCC) in a graph of particles.
	Two nodes (i.e., particles) in the graph are connected if their metric is less than a predefined linking length.
	Finding the exact FoF groups requires searching all nodes in the graph and computing the distances between the nodes.
	
	Distance calculations among nodes (particles) can be minimized using a tree structure.
	The most commonly used is a KD tree structure \citep[][]{Bentley75, Moore01}.
	The KD tree is built by iteratively dividing a domain (parent node) into the two subdomains (child nodes), starting from the entire volume.
	A child node is no longer split up if the number of particles in the node is less than a predefined size\footnote{32 in this study}, called a leaf node.
	All the other nodes are then dubbed a split node.
	In each division, a value (split value) at a dimension (split dimension) cuts a parent node into the two child nodes: a left child node has particles whose coordinates at the split dimension are less than the split value, and the right child node possesses the other particles.
	In common, a dimension along which a node has the maximum length, and a value with which the two child nodes have the same number of particles, are used to construct a KD tree.
	
	Searching for complete FoF groups with the KD tree is usually performed in a linked-list representation scheme \citep[e.g.,][]{Cormen09} combined with the depth-first or breadth-first tree search because of its simplicity.
	Figure \ref{fig:fig LLR} illustrates how the linked-list representation method finds an MCC.
\begin{itemize}
	\item[$-$] The first particle in the list is tagged as a head (blue arrow), and a particle at the end of the list is tagged as a tail (red arrow).
	
	\item[$-$] (\textit{Step 1}) Add particles to the list whose distance to the head particle is less than the linking length (white circles).
	If there is a newly linked particle, the tail tag moves to the last particle.
	
	\item[$-$] (\textit{Step 2} - \textit{Step N}) The next particle of the previous head particle becomes the new head particle.
	Repeat \textit{Step 1} for the new head particle.
	If there is no newly linked particle, the tail particle does not change.
	
	\item[$-$] If the new head particle is the same as the tail particle, stop the process.
\end{itemize}
	This process finds the exact list of particles of an FoF group for a given set of particles and the linking length.

	Each step of the linked-list representation method requires calculating the distances between adjacent particles and the head particle.
	Without any secondary implementation, this is a very time-consuming calculation, scaled by $O(N^2)$, $N$ being the number of particles.
	Conversely, using a tree structure such as the KD tree can significantly reduce the number of computations, which is scaled by $O(N \log N)$ in an ideal case.
	This is because the binary search with the KD tree allows one to find a set of adjacent particles inexpensively.
	
	However, if the linking length is much larger than the inter-particle distances, too many KD tree nodes are selected by the binary search.
	This results in a large overhead of distance calculations.
	This often happens in the over-dense regions of FoF groups as already reported by many studies \citep[e.g.,][]{Behroozi13,FM17}.
	Therefore, the fraction of the linking length to the inter-particle distances is a deciding factor of the FoF efficiency with the KD tree.
	For example, in Figure \ref{fig:fig ltest}, we perform 3D FoF searches with different values of linking length: from 0.0001 to 0.5 times to the mean inter-particle distance of dark matter particles where $\times0.2$ is the reference value.
	We multiply a constant value to each test result for illustrative purposes, by which wall-clock time values in each test have vertical offset in the plot. 
	For the test runs with shorter linking length ($\times 0.0001 - \times 0.0005$), 3D FoF searches follow the ideal scaling ($\propto N \log N$).
	In contrast, some 3D FoF groups, usually from the most massive ones, deviate from the ideal scaling and show the poor scaling ($\propto N^2$) at the runs with longer linking length.
	Therefore, as a simulation has higher particle and spatial resolutions, the inter-particle distances at the central regions of FoF groups get smaller compared to the fixed linking length, resulting in a poor performance of FoF searches.
	
	Figure \ref{fig:fig GalandNode} visually illustrates the above issue.
	Panel (a) of Figure \ref{fig:fig GalandNode} shows the projected mass density of the most massive 3D FoF group, containing $4,166,225$ particles at $a = 0.289$ in the \NH\ simulation.
	The central area of the group (white square in Panel (a)), a box of $6\,{\rm kpc}$ in length, is enlarged in Panel (b), which is made up of $2,977,072$ particles. 
	More than $70\,\%$ of the total particles are located in the central region.
	Panel (c) displays the same plot as Panel (b), with leaf nodes of the KD tree drawn together.
	The white-colored lines in Panel (c) represent the boundaries of the leaf nodes of the KD tree in a thin plane ($\Delta {\rm Z} = 0.06\,{\rm kpc}$).
	The search sphere of the radius with the linking length ($1.998\,\rm{kpc}$) is indicated by the yellow circle.
	In the search sphere, there are $1,594,194$ particles, corresponding to $72,087$ leaf nodes.
	A large number of particles (leaf nodes) within a single search radius imply that a standard 3D FoF implementation will require a large amount of computation.
	
	The number of calculations required to find (3D) FoF groups can be approximated using a more analytical approach.
	For an $i$-th target particle, the neighboring particles selected by the binary search are checked to determine whether they are indeed within a search sphere.
	In the case where a simulation has poor particle resolution, the linking length is much shorter than the inter-particle distance of stellar particles.
	Then, the selected neighboring particles through the binary search locate at a few leaf nodes, and thus the number of distance calculations for these particles ($N_{\rm cal, i}$) is approximately the number of particles within a leaf node---a few tens in common.
	In the other case of a simulation with high particle resolution, the selected neighboring particles are distributed in many leaf nodes.
	Thus, $N_{\rm cal, i}$ is equal to the number of particles in those leaf nodes and should be greater than the number of particles within the search sphere centered at the $i$-th particle:
	\begin{equation}
		N_{\rm cal, i} \ge n_{\rm i}\,\frac{4\pi}{3}\,l^3,
	\end{equation}
	where $n_{\rm i}$ is the number density at the location of the $i$-th particle, and $l$ is the linking length.
	The equality is satisfied if the number of particles in a leaf node is sufficiently small.
	If an FoF group consists of $N_{\rm p}$ particles, then the total number of calculations required for identification ($N_{\rm cal}$) can be approximated as
	\begin{equation}
	\begin{aligned}
		N_{\rm cal} &\ge \sum_{i=1}^{N{\rm p}} n_{\rm i} \frac{4\pi}{3} l^3 \\
		&= \frac{4\pi}{3 m_{\rm p}^2} l^3 \int_{0}^{R} 4 \pi r^2 dr \rho^2 (r).
	\end{aligned}
	\end{equation}
	The latter equality is derived by assuming the following.
	(i) All particles have the same mass ($m_{\rm p}$) and are distributed within the sphere of radius $R$.
	(ii) The number of particles is sufficiently large and the radius of the sphere is much larger than the linking length, to treat the summation term as an integral.
	(iii) The density distribution ($\rho$) is spherically symmetric, that is, $\rho = \rho(r)$.
	In many potential models (e.g., NFW, Plummer), the value of the integral term can be expressed as the central density term ($\rho_{\rm 0}$) and the characteristic scale radius ($a$): $\rho_{\rm 0}^2 / {a^3}$.
	Therefore, the number of calculations is only dependent on the central density, i.e.,
	\begin{equation}
	\label{eq:numberofloops}
		N_{\rm cal} \propto \rho_{\rm 0}^2.
	\end{equation}
	This highlights the issue of the poor performance of the FoF galaxy finding in an over-dense region because the central density of galaxies is very high.
	We also confirm the linear scaling between the central density and total stellar mass of the galaxies in the \NH\ simulation data.
	
	Therefore, in the \NH\ case in which the linking length is much larger than the inter-particle distance of {\em stellar particles}, the central number density---particle resolution--- is a deciding factor of the efficiency of the FoF.
	Considering the relationship between the central density and total stellar mass of the galaxies, galaxies containing more particles will take longer to be searched.
	To test this, we run the original version of \VR\ on the \NH\ dataset at a snapshot of $a = 0.289$ with $47,218,669$ stellar particles in the zoomed volume.
	The snapshot is chosen considering that the run must be completed within a reasonable human time while allowing several normal galaxies to form.
	We also provide a run report using the snapshot at $a = 0.461$ in Appendix \ref{sec:Appendix}.
	
\begin{figure}
\centering
\includegraphics[width=0.45\textwidth]{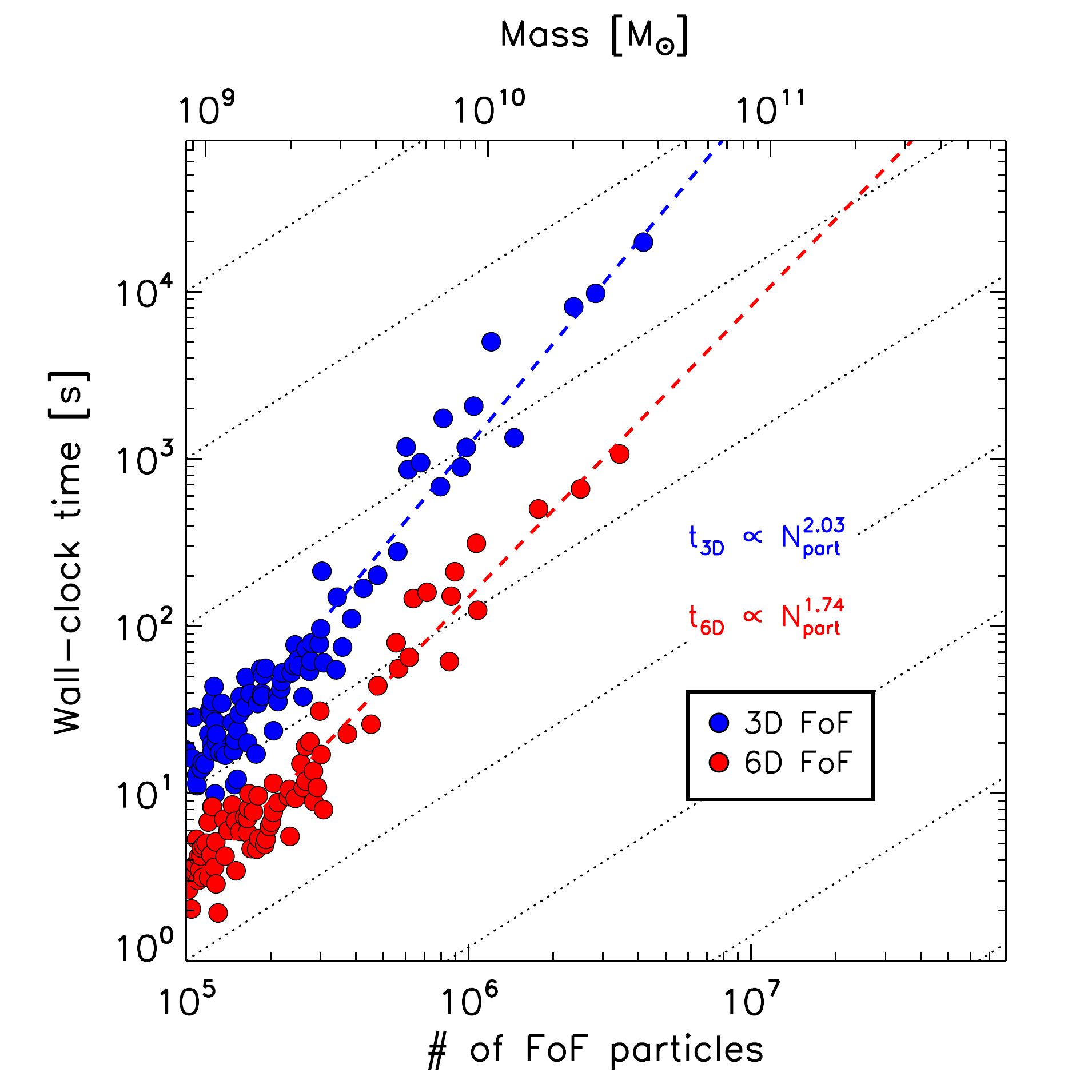}
\caption{Wall-clock times versus the number of particles in each FoF group using the {\em original version} of \VR.
Stellar mass corresponding to the number of particles is shown on the top axis.
Wall-clock times are measured for 3D FoF (blue circle) and 6D FoF searches (red circle).
The dashed lines represent the fitting curves of the wall-clock times, and the dotted lines are the ideal scaling relation ($\propto N_{\rm part} \log{N_{\rm part}}$), showing a very poor scaling of the original runs ($\propto N_{\rm part}^{2.03}$ for the 3D FoF search and $\propto N_{\rm part}^{1.74}$ for the 6D FoF search).
The largest groups take very long time to be searched ($> 10^{4}$ seconds for the 3D case and $> 10^{3}$ seconds for the 6D case).}
\label{fig:fig Runtime_All}
\end{figure}

\begin{figure*}
\centering
\includegraphics[width=0.95\textwidth]{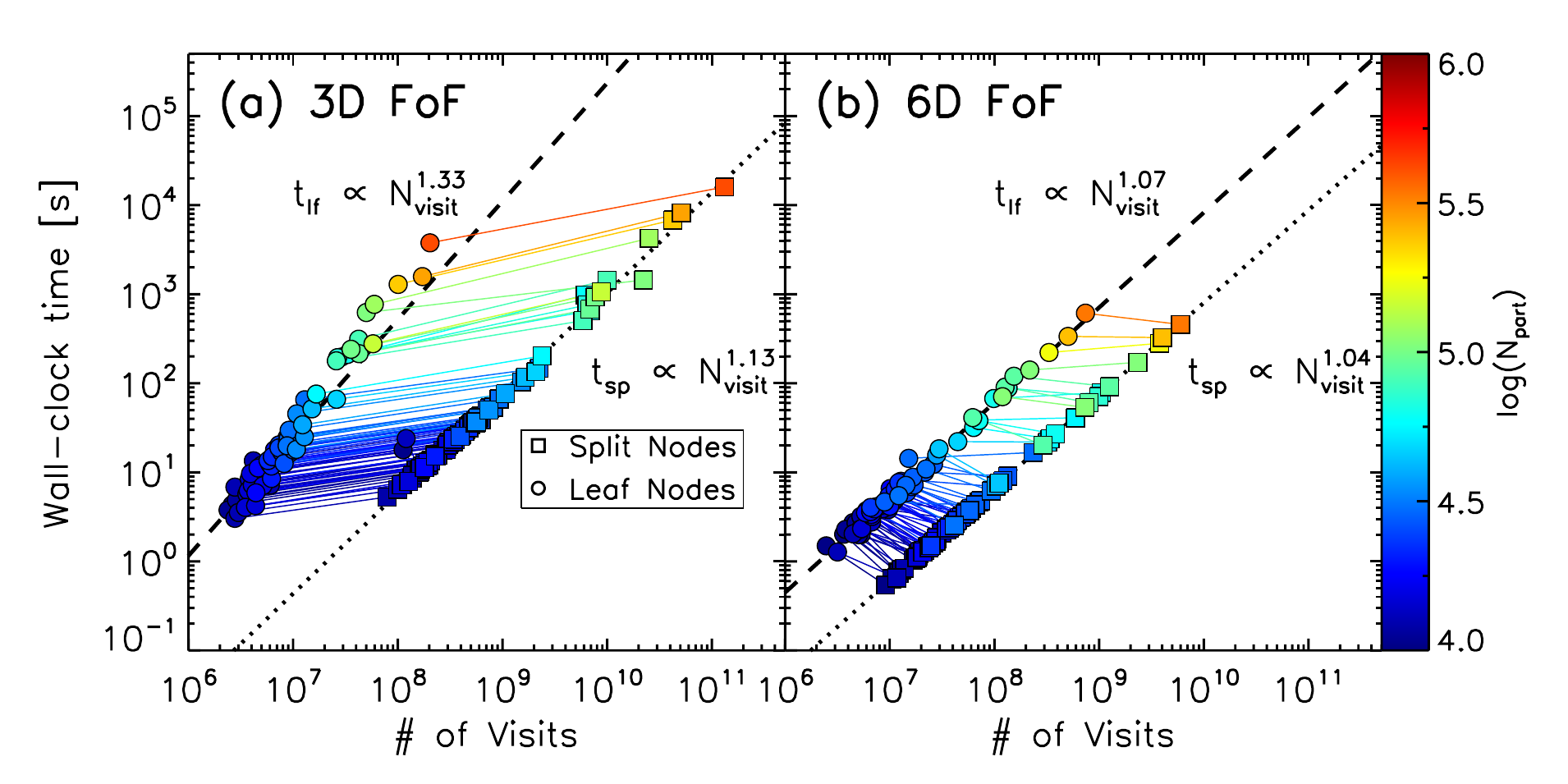}
\caption{Total wall-clock times of FoF searches on split and leaf nodes as a function of the number of visits to nodes.
Panels (a) and (b) illustrate the results from the 3D FoF and 6D FoF searches, respectively.
The wall-clock times of the split (leaf) nodes are represented by squares (circles).
The color of the data points shows the number of particles in the FoF groups, and the data points of the same FoF group are connected to the same colored line.
The dotted and dashed lines represent the fitting curves of the data points of the split nodes and leaf nodes, respectively.
All four relationships show a tight correlation, implying that the number of visits is a key factor in the performance of the FoF search.}
\label{fig:fig Vtimes_All}
\end{figure*}

	During the run, we measured the wall-clock time for finding the FoF groups.
	\VR\ searches 3D FoF groups first, followed by 6D FoF groups inside each 3D FoF group, so the wall-clock time measurement is separated for each type of search\footnote{The actual run of \VR\ recursively searches substructures within the found 6D FoF groups.
	However, we do not include this process because we focus only on the FoF search performance.}.
	Figure \ref{fig:fig Runtime_All} shows the results of wall-clock times as a function of the number of FoF particles using {\em the original \VR\ code}.
	The stellar mass corresponding to the number of particles is plotted on the top horizontal axis.
	Wall-clock times in 3D FoF and 6D FoF searches are marked with blue and red circles, respectively.
	The background dotted lines represent the ideal scaling relationship when using a tree structure ($\propto N\log{N}$).
	First, the figure shows that the original version of \VR\ performed very poorly for the \NH\ simulation dataset, taking more than $10,000$ seconds to find the largest 3D FoF group.
	The 6D FoF search has a slightly better speed because of the smaller linking length used in the 6D search, but it still shows a similar slope for the wall-clock time as for the 3D FoF case.
	The scaling is approximately $\propto N^{2.03}$ for the 3D FoF case and $\propto N^{1.74}$ for the 6D case, which is similar to the expected relationship in Equation \ref{eq:numberofloops}.
	The overall scaling implies that it will take a longer physical time to run for a snapshot that includes more massive galaxies.
	For example, another test run using a snapshot with a later epoch ($a = 0.461$) takes more than a week for execution (see Appendix \ref{sec:Appendix} for details).
	
	To study the issue with the current FoF implementation, we examine the number of visits to tree nodes during a search.
	If the visited node is fully-connected, i.e., all particles in the node are linked to FoF groups, there is nothing to calculate, so this visit is not counted.
	When a large FoF group is made up of a large number of tree nodes, the total number of visits to this FoF group should be large.
	The relationship between the wall-clock time and the number of visits is likely to be linear because the same number of calculations will be performed on each node visit.
		
	Figure \ref{fig:fig Vtimes_All} shows wall-clock times as a function of the number of visits again using {\em the original \VR\ code}.
	Panels (a) and (b) in the figure show the wall-clock times of the 3D FoF and 6D FoF groups, respectively.
	In each panel, the wall-clock times on split ($t_{\rm sp}$) and leaf nodes ($t_{\rm lf}$) for FoF searches are marked with colored square and circular data points, respectively, and the color scheme shows the number of FoF particles.
	In principle, the total wall-clock time for searching an FoF group is the summation of $t_{\rm lf}$ and $t_{\rm sp}$ that are connected with the same colored solid line.
	The fitting functions written in each panel are drawn with black dashed lines (for leaf nodes) and dotted lines (for split nodes). 
	All four relationships have a power index close to 1, which is consistent with the expected linear correlation between the wall-clock time and the number of visits\footnote{We note that the wall-clock times shown here include the time for report-related processes and are, therefore, the upper limits. The actual wall-clock times are lower by approximately $20 \%$ or less.}.
			
	For a fixed number of visits, the time spent on the leaf nodes is much longer than that for the split nodes.
	This is simply because, in \VR, the full distance calculation between particles is performed only at the leaf nodes.
	
	Considering the apparent tightness of the relationships, the number of visits to nodes is likely to be a key factor in the efficiency of FoF searches.
	Therefore, {\em our strategy to increase the efficiency of the FoF search described in the following section involves reducing the number of node visits for a given FoF group}.

\section{Implementations}
\label{sec:Imple}

	Considering that the search performance (in terms of computing time) strongly depends on the number of visits to tree nodes, we aim to minimize the number of visits to tree nodes, especially in dense regions, by devising additional algorithms that determine the number of visits and modifying the KD tree structure.

\subsection{Node-Closure}
\label{sec:Imple-NC}

	In the KD tree method, a neighbor query to search for particles in the vicinity of the target particle is based on a binary search scheme.
	Specifically, from the root node, a child node is visited if it contains the target particle or is sufficiently close to the target particle.
	Only when a leaf node is visited are distance calculations performed.
 	This is how the use of KD tree makes the neighbor search inexpensive.
	However, the advantage of the tree structure is lost in over-dense regions because there are too many leaf nodes in the search window, as addressed in Section \ref{sec:Background}. 
		
	To increase the efficiency of the binary search in over-dense regions, the following conditions (namely, ``Node-Closure'') have been employed.
\renewcommand{\labelenumi}{\roman{enumi})}
\begin{enumerate}
	\item If a node is tagged as closed, that node is no longer visited.
	
	\item When a leaf node is fully-connected, i.e., all particles within the node are identified as members of FoF groups, then the leaf node is closed.
	
	\item When the two child nodes of the same parent node are closed, the parent node is also closed.
\end{enumerate}
    Under these conditions, it is expected that many nodes in dense regions will be closed after a single neighbor query, reducing the depth of the KD tree structure.
    This methodology will continuously reduce the number of active nodes that must be visited, which results in duplicate removal of fully-connected nodes from the active search.
    For example, after finding all the neighboring (within a linking length) particles for the centered particle in Panel (c) of Figure \ref{fig:fig GalandNode}, all their corresponding leaf nodes and split nodes are ``closed'', such that the original 72,087 actively searched nodes is reduced to 41 open nodes, which happen to reside near the edge of the search window.
        The use of Node-Closure has been associated with good performance gains \citep[see e.g.,][]{Kwon10, FM17, Springel21}.
	The original version of \VR\ also utilizes the first two conditions above at the {\em leaf node level}.
	In this study, we implement the Node-Closure scheme for the {\em split nodes} (the third condition) to achieve further improvement, as presented in Section \ref{sec:Results}.

\subsection{Node-Geometry}
\label{sec:Imple-NG}

	The next implementation is regulating node visits by considering the geometry between a node and a target particle.
	If a visited node is fully inside or outside the search sphere of the target particle, there is no need to visit its child nodes any further.
	The number of visits to nodes gets reduced this way by skipping unnecessary visits, thereby increasing the efficiency of the FoF search.
	However, because the shape of a node and the distribution of particles within it are arbitrary, it is non-trivial to determine unnecessary visits without spending a large amount of computing time.
	We introduce a simple algorithm for this purpose.

\begin{figure}
\centering
\includegraphics[width=0.45\textwidth]{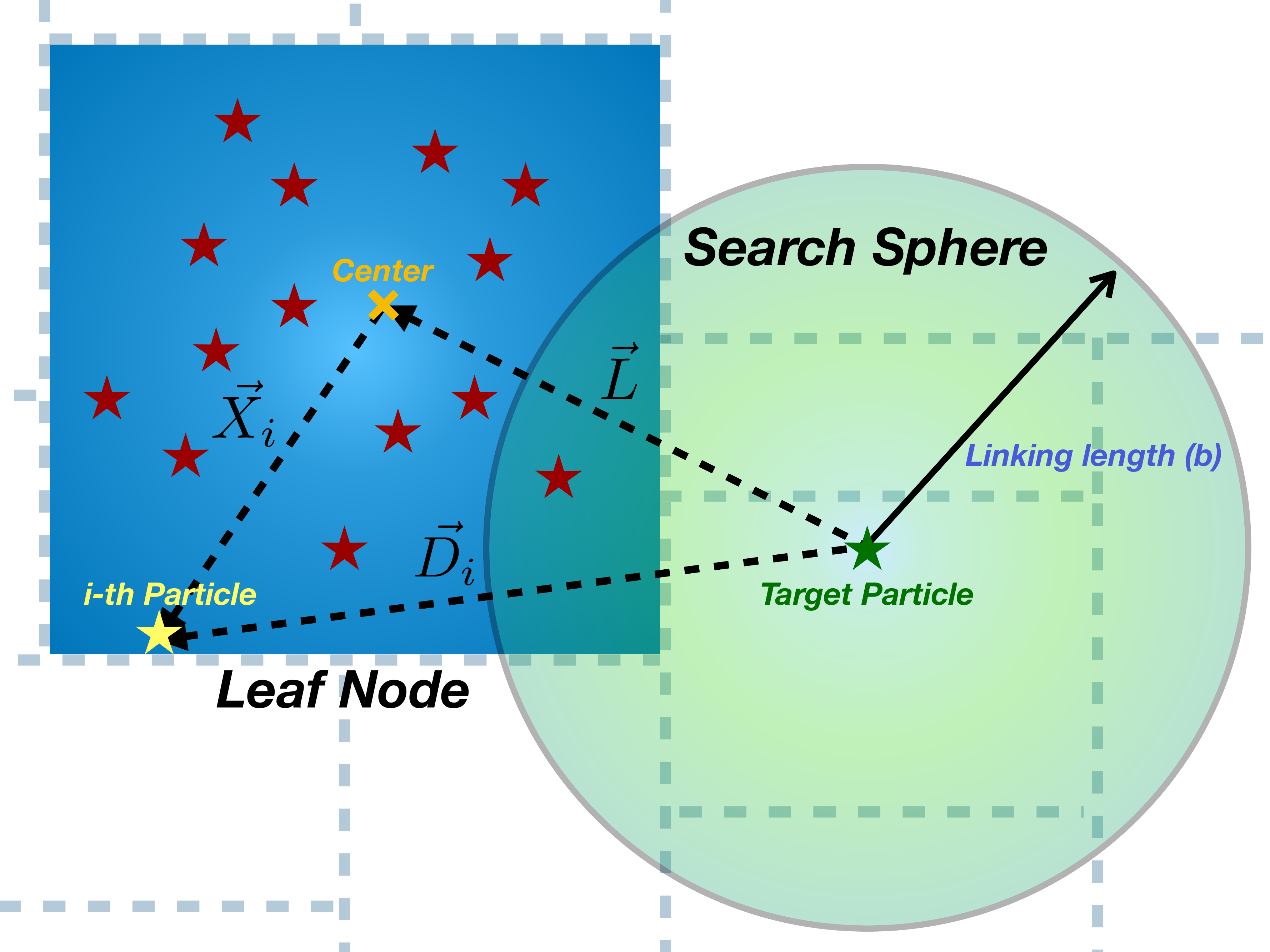}
\caption{Schematic diagram of the Node-Geometry scheme.
A target particle and the search sphere of the target particle whose radius is the linking length ($b$) are drawn with the green star symbol and the green-shaded sphere, respectively.
A visited leaf node is shown with a blue box, and the particles inside the node are illustrated by red and yellow star symbols.
The orange $\times$ symbol denotes the center of the node.
The arrows indicated with $\vec{X_{i}}$, $\vec{D_{i}}$, and $\vec{L}$ correspond to positional vectors of the $i$-th particle with respect to the center of the node, $i$-th particle with respect to the target particle, and the center of the node with respect to the target particle, respectively.}
\label{fig:fig NodeSkipSchematic}
\end{figure}

	Figure \ref{fig:fig NodeSkipSchematic} shows a schematic diagram of the algorithm.
	A target particle is indicated by a green-colored star symbol.
	Its search sphere, whose radius is the linking length ($b$), is represented as a green-shaded sphere.
	The blue square is a visited node, and the particles in it are indicated by red and yellow star symbols.
	The center of the node is denoted by the orange $\times$ symbol.
	The positional vectors $\vec{D_{i}}$, $\vec{X_{\rm i}}$, and $\vec{L}$ correspond to each side of a triangle defined by the center of the target particle (green star symbol), the center of the node (orange $\times$ symbol), and the $i$-th particle in the node (yellow star symbol).
	Given the geometry, this node can be fully-connected to the target particle if ${\rm max}\{\|\vec{D_{i}}\|\}\,<\,b$ and can be skipped to visit if ${\rm min}\{\|\vec{D_{i}}\|\}\,>\,b$.
	Therefore, the key to determining further visits is on the algorithm how to determine the values of ${\rm max}\{\|\vec{D_{i}}\|\}$ and ${\rm min}\{\|\vec{D_{i}}\|\}$ for each node visit.
	A triangle whose lengths are $||\vec{L}||,\ {\rm max}\{||\vec{X_{\rm i}}||\},\ {\rm and}\ ||\vec{D_{\rm i}}||$ gives the trigonometric inequality:
	\begin{equation}
	\label{eq:node geometry}
		| \, \|\vec{L}\| - {\rm max}\{ \| \vec{X_{i}} \| \} \, | \, \leq \, \| \vec{D_{i}} \| \, \leq \, \| \vec{L} \| + {\rm max}\{ \| \vec{X_{i}} \| \}.
	\end{equation}
	Each equality is satisfied when two spheres, the search sphere and the sphere centered at the node center with radius ${\rm max}\{ \| \vec{X_{i}} \| \}$, are inscribed (left equality) or circumscribed (right equality).
	Then, the remaining work defines the center of the node and computes ${\rm max}\{ \| \vec{X_{i}} \| \}$ for the center.
	
	The center of the node can be defined in various ways.
	A simple and effective method is to minimize the norm of the residual $f_{i}$, defined as
	\begin{equation}
		f_{i} = (\, \| \vec{L} \| + \| \vec{X}_{i} \| \,) - \| \vec{D_{i}}\|.
	\end{equation}
	There are several methods for evaluating the norm of residuals, such as $\sum \| f_{i} \|_{2}$ and ${\rm max}\| f_{i} \|_{1}$, and finding the smallest sphere surrounding particles in the node can be an example.
	We decide to define the center as the mean of the positional (and velocity in the 6D case) vectors to minimize the computational time when constructing a tree.
	In the actual implementation, when building a KD tree, the center of each node and ${\rm max}\{ \| \vec{X_{i}} \| \}$ are calculated and stored at the pointer of the node.
	When performing the FoF search in each node visit, we load the values and determine the lower and upper limits of $\| \vec{D_{i}} \|$ based on Equation \ref{eq:node geometry}.
	If the lower limit is greater than the linking length, then visiting to the node is skipped. In the similar manner, if the upper limit is smaller than the linking length, the node is opened.
	If ${\rm max}\{ \| \vec{X_{i}} \| \} < b$ is satisfied for an opened node, then the node can be fully-connected, i.e., distance calculations for this node are skipped.
	
	This algorithm is effective as long as each node has a sufficiently large number of particles, which is the case in this study.
	Besides, it has the advantage of taking into account the distribution of particles within the node only and not the full geometric shape of the node. 
	
\subsection{Splay operation}
\label{sec:Imple-Splay}

	The third implementation is to modulate the order of neighbor queries of the target particles in the linked-list shown in Figure \ref{fig:fig LLR}.
	In the ``linked-list representation method'', finding an FoF group follows two steps: 1) finding all the member particles of the same group as the target particle and 2) confirming that there are no more linked particles outside the group. 
	Finding all member particles in the first step coincides with enlarging the volume retrieved by the search spheres until the FoF volume is completely covered.
	For example, the neighbor query of the first target particle finds particles within the search sphere.
	Then, all neighbor queries performed on the added particles in the first neighbor query identify member particles at a distance $b$ to $2b$ from the first target particle, $b$ being the linking length.
	As this step is extended further out from the first target particle, the volume containing all member particles becomes progressively larger.
	After finding all the member particles, we confirm that there is no further particle outside this volume to be added.
	
	We attempt to improve the FoF search process by minimally overlapping search spheres.
	In the KD tree implementation, the next target particle is likely to be one of the particles close to the current target particle.
	This means that, in the next neighbor query, there are too many already-connected particles to be searched.
	In other words, the two consecutive search spheres overlap too much.
	This is likely to decrease the efficiency of FoF searches with the KD tree if the leaf node size is much smaller than the search sphere because too many nodes should be repeatedly visited, which is the case for \NH\ (e.g., Panel (c) of Figure \ref{fig:fig GalandNode}).
	Therfore, the ideal situation would be to select the particle at the furthest distance from the target particle as the next target particle.
	This would result in two consecutive search spheres overlapping only up to $31.25 \%$ of the search sphere volume\footnote{The value is obtained from the overlapping volume of the two identical 3D spheres when they are at the same distance apart as the radius.}.
	
	Choosing one of the particles farther from the target particle as the next head particle can be challenging; therefore, we introduce the following simple implementation.
	In a KD tree binary search, the neighbor query starts at the root node and recursively visits the child nodes.
	For each node visiting, the child node is visited if it contains the target particle (\textit{Case 1}) or is sufficiently close to the target particle (\textit{Case 2}).
	In the implementation of the algorithm, we set the child node with \textit{Case 1} to always visit before the child node corresponding to \textit{Case 2}.
	This implementation naturally causes the last added particle to be further than the neighboring particles. 
	The splay operation, therefore, is defined as having the last added particle as the next target particle after every neighbor query.
	
	The splay operation expands the volume of the member particles efficiently but does not improve the run speed by itself.
	This is because the splay operation only changes the order of the target particles but not the actual number of visits to nodes.
	Only when combined with the Node-Closure and/or Node-Geometry scheme does the Splay operation show a significant improvement in performance.

\subsection{KD tree structure}
\label{sec:Imple-Tree}

	The last implementation concerns the structure of the KD tree.
	The purpose of modifying the tree structure is to consider particle clustering as much as possible before performing the FoF search.
	This way, we expect each node to be more fully-connected, thus increasing the efficiency of the FoF search.
	As described in Section \ref{sec:Background}, the criteria that divide a node into two child nodes, that is, the split dimension and split value, determine the structural shape of the KD tree.
	The method commonly used to select a split dimension is the axis whose length of a node is the longest, which produces leaf nodes that are fairly uniform in shape.
	The median value of the particle positions along the split dimension is usually defined as the split value.
	Both lead to a balanced KD tree, with which all leaf nodes have a similar number of particles.

	We add two options for determining the split dimension.
	The first option is to select the dimension with the largest dispersion of the coordinate components of the particles.
	The second is to choose the dimension along which the Shannon entropy is maximized \citep[][]{Elahi19a}.
	
	Regarding the split value, we also add one more option beyond employing the median value: splitting where the distance in the split dimension between the two closest member particles is the greatest.
	This sometimes results in very unbalanced trees, therefore, we search for the split position using only the inner 75\% of the member particles, excluding the others on either end.
	This way, each child node defined in the very early stage of tree making likely contains particles that will eventually belong to a separate FoF group (e.g., a galaxy).
	We note that this choice of split value also provides a better option for decomposing domains for parallelization, which avoids expensive calculations when stitching FoF groups after the local search (see Section \ref{sec:Results-Tree}).

\begin{figure*}
\centering
\includegraphics[width=0.95\textwidth]{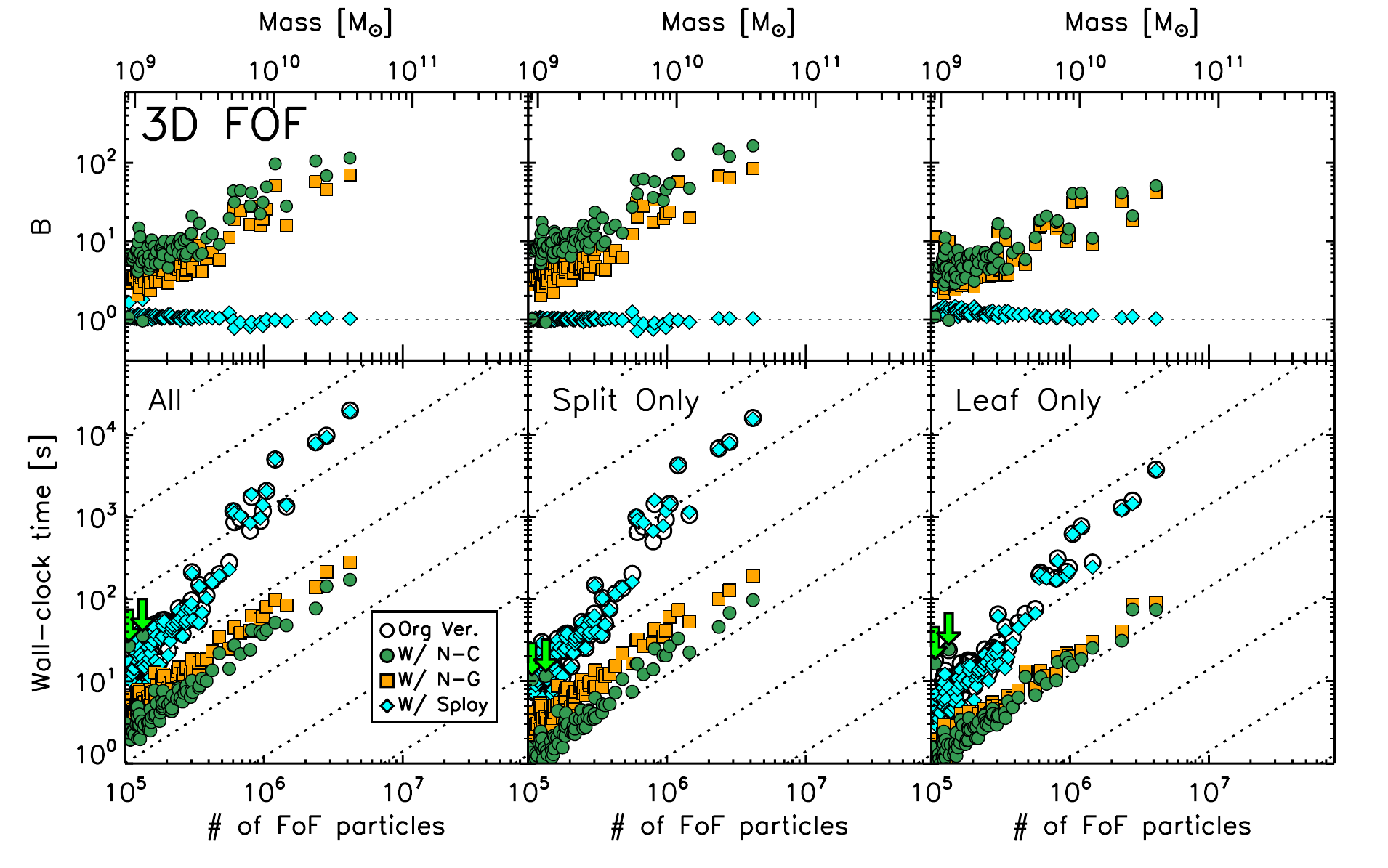}
\caption{Wall-clock times of the 3D FoF search (bottom panels) and the \BF\ parameter of the groups (upper panels) as a function of the number of particles.
The stellar mass corresponding to the particle number is displayed on the top horizontal axis.
The colored symbols in the bottom panels indicate the results of the individual application of each implementation (see the legend), and the open circles represent the original wall-clock times, as shown in Figure \ref{fig:fig Runtime_All}.
From left to right, each column shows the total time (left), time spent on split nodes (middle), and time spent on leaf nodes (right), respectively.
The background dotted lines are the guidelines for ideal scaling ($\propto N\,\log N$).
The two green arrowed symbols in the bottom panels are the two 3D FoF groups that show poor performance with the Node-Closure scheme (see text for details).
Although it is not the case for Splay, the other two implementations show significant performance gains.}
\label{fig:fig Runtime_3D}
\end{figure*}


\section[]{Results}
\label{sec:Results}

In this section, we show the speed improvement attributed to the application of the implementations discussed in Section \ref{sec:Imple}.
We note that {\em all the test runs with different implementations find the same exact FoF groups compared with the original \VR\ run because the FoF groups are unique for a given linking length.}
We note that all tests shown through this section are carried out on the same machine with the same compiler and the same optimization.
Moreover, we confirm that there is no or little extra usage of memory with the implementations, and hence, we use the change of wall-clock time of FoF searches as our main indicator of the performance variation.
In Sections \ref{sec:Results-3DFoF} and \ref{sec:Results-6DFoF}, we focus on the results of 3D and 6D FoF performance, respectively, with Node-Closure (N-C), Node-Geometry (N-G), and Splay.
In Section \ref{sec:Results-Tree}, we address the performance improvement by modulating the KD tree structure.

\subsection[]{3D FoF}
\label{sec:Results-3DFoF}

	The first results correspond to the performance gains achieved by applying N-C, N-G, and Splay to the 3D FoF search.
	Figure \ref{fig:fig Runtime_3D} shows the 3D FoF performance results obtained with only one of the three implementations.
	In the bottom panels, the symbols represent the wall-clock times of the 3D FoF search as a function of the FoF group size in the particle number.
	Open circles are the results from the original run, as shown in Figure \ref{fig:fig Runtime_All}.
	Green circles, orange squares, and cyan diamonds are the results of applying N-C, N-G, and Splay, respectively.
	The three columns show the result of total wall-clock time (left), time spent on split nodes (middle), and time spent on leaf nodes (right).
	The background dashed line is the ideal scaling ($\propto N\log{N}$).
	The upper panels show the ratio of the wall-clock time before and after the implementation of the groups: $\BF\, = 1$, meaning no change and $\BF\, = 1,000$ meaning 1,000 times faster than before.
	
	The original \VR\ points lie behind the Splay points as splay does not drastically change the time taken.
 	This means that applying the Splay operation alone does not significantly improve the efficiency of the 3D FoF search.
	As we argued in Section \ref{sec:Imple-Splay}, this is because the Splay operation only modulates the order of a linked-list and does not affect the number of visits to nodes.
	In contrast, the use of N-C (green circles) or N-G (blue squares) significantly improves the efficiency of the 3D FoF search.
	For example, the most massive FoF group takes only $0.65 \%$ ($\BF\, = 154$) and $1.07 \%$ ($\BF\,= 93$) of the wall-clock times with N-C and N-G, respectively, compared with the original run.
	We confirm that this performance enhancement is achieved primarily through the reduction of the number of visits to nodes and conclude that N-C and N-G are both effective for our purpose.

	We further elaborate on how exactly the use of N-C reduces the number of visits.
	Using the N-C scheme, all leaf nodes and their parent nodes in the search sphere are \textit{closed} for visits once the neighbor query of a target particle is performed, reducing the depth of the tree structure.
	All 3D FoF groups presented in Figure \ref{fig:fig Runtime_3D} have 95 \% of their nodes closed.
	
	Regarding the N-G scheme, this method reduces the active nodes effectively reducing the depth of the tree searched for a given particle.
	During the FoF search, if the upper limit of Equation \ref{eq:node geometry} is less than the linking length on a visited node, the node visit ends at that node, and no more child nodes of this node are visited.
	In practice, approximately $30-80\,\%$ of visits to split nodes are skipped this way during the 3D FoF search.
	Conversely, if the lower bound of Equation \ref{eq:node geometry} is greater than the linking length, the algorithm blocks the visit to the child nodes in the same manner.
	However, the latter case rarely occurs, by which about $0.2\,\%$ of split node visits are skipped.
	Thus, both N-C and N-G schemes improve the FoF search efficiency in a manner similar to shallowing the depth of the KD tree structure.
		
	The time taken on the leaf nodes is also substantially reduced with N-C and N-G.
	However, considering that both approaches were already implemented in the original version of \VR\ at the {\em leaf node level}, this reduction is rather unexpected.
	The improvement occurs because, in our definition, the wall-clock time on leaf nodes includes that of checking process of whether to visit a leaf node.
	In both implementations, node visits are blocked from the upper split nodes, and thus the number of checking processes for leaf nodes is also reduced.
	Therefore, the improvement seen in the leaf nodes is a byproduct of fewer visits to split nodes and also of the definition of wall-clock time on leaf nodes.
	
	The test run with N-C (green circles) is generally faster than that with N-G (orange squares).
	In fact, the number of node visits for each FoF group is similar between them, but the computation time is faster with N-C.
	This is because of the additional calculations that are used for computing the limits of Equation \ref{eq:node geometry} in the N-G scheme every time a node is visited.
	Because the number of visits to nodes is already high, such a minor additional calculation can easily add up to be noticeable.

	However, in the N-C scheme, two 3D FoF groups are found with no performance gain (two arrowed green circles in the bottom panels of Figure \ref{fig:fig Runtime_3D}).
	The two groups actually consist of multiple substructures connected to each other via particle bridges.
	The shape of these structures can adversely affect the rate of node closure.
	For example, we confirm that the nodes in ordinary FoF groups are almost $100\,\%$ closed after the FoF search, whereas only a few nodes are closed for the two groups.
	Both groups also show unusually high numbers of visits (by more than a factor of 10) compared with the other 3D FoF groups\footnote{In fact, these two groups are the two outliers at roughly ($10^8$, 20) in Figure \ref{fig:fig Vtimes_All}-(a).}.
	The same situation occurs in another snapshot (see Figure \ref{fig:fig Runtime_3D_400}), implying that the implementation of N-C alone is not the best choice.
	Indeed, these two groups show a good performance with the N-G scheme.

\begin{figure}
\includegraphics[width=0.45\textwidth]{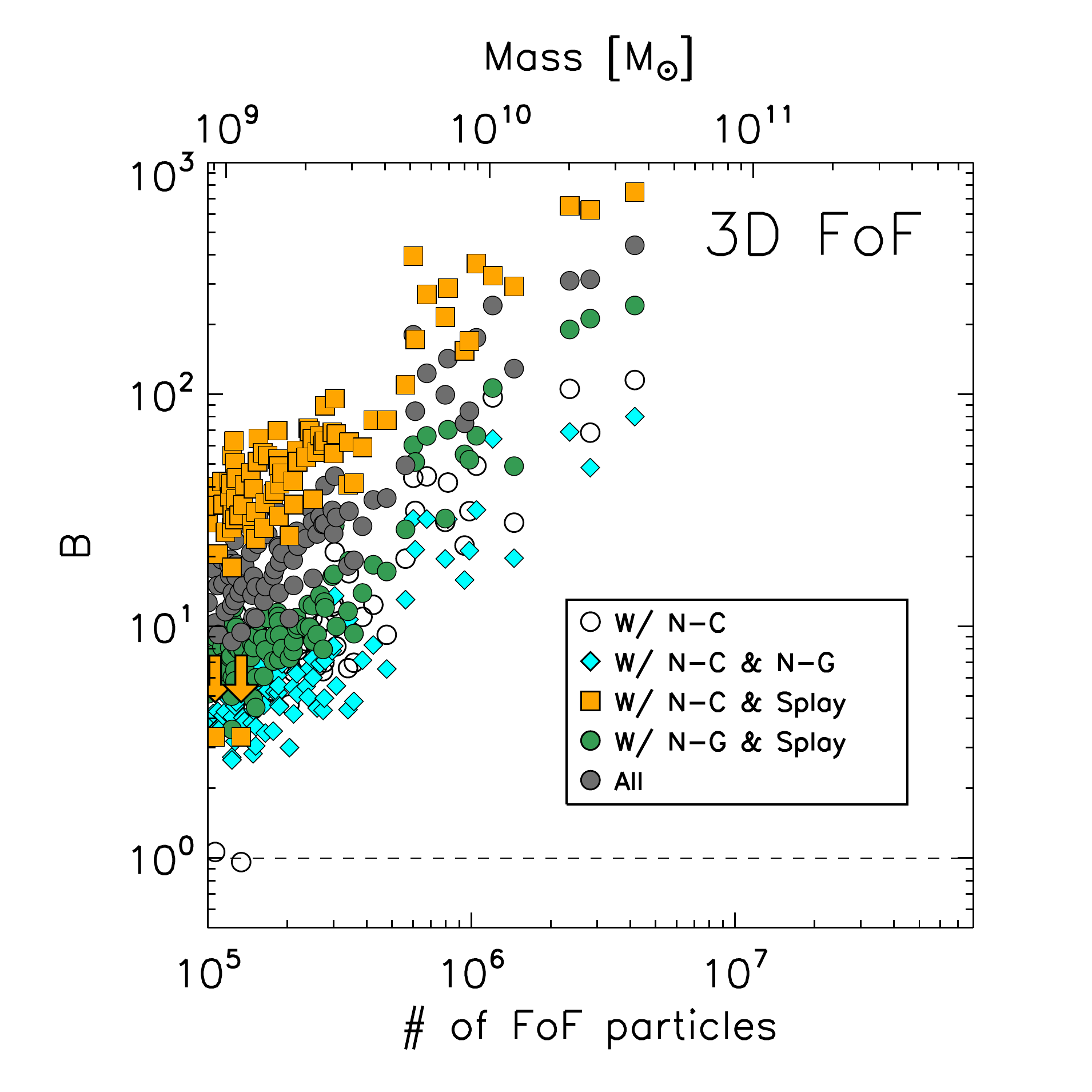}
\caption{\BF\ parameter ($\BF\,= 1,000$ means 1,000 times faster) of the 3D FoF groups for each combination of our implementations.
For comparison, the results with N-C showing the best performance of single implementations are represented by open circles.
The results for each combination (see the legend for details) are denoted by different colors and symbols.
The combination of N-C and Splay shows the best performance.
However, groups with relatively low performance improvements appear (two arrowed orange squares with $\BF\,\sim 3$), therefore, it is recommended to use all implementations for the 3D FoF search.}
\label{fig:fig Runtime_comb_3D}
\end{figure}

	Finally, the effect of combining all implementations is shown in Figure \ref{fig:fig Runtime_comb_3D}.
	The figure shows the \BF\ parameter defined earlier (the same format as the upper panels of Figure \ref{fig:fig Runtime_3D}).
	For a comparison with the results with a single implementation, the results of N-C, which shows the best performance among the single implementations, are shown (open circles).
	All combinations show very impressive performance improvements, in which the wall-clock time fractions of the most massive 3D FoF group are greater than 100.
	However, applying both N-C and N-G schemes (cyan diamonds) show poorer results than the single implementation of N-C.
	Since each implementation already effectively reduces the number of visits, applying both schemes only results in more computations without any benefits, causing poorer performance.
	The combination of N-C and Splay operations (orange squares) gives the best result, with which the wall-clock time of the most massive 3D FoF group is $1,000$ times smaller than the original value.
	Because finding the most massive FoF group dominates the total wall-clock time of the 3D FoF search, we note that the speed-up of the most massive FoF can be directly indicative of the overall speed-up.
	However, as shown in the single implementation of N-C, the two 3D FoF groups show relatively lower performance improvements compared with the other groups, which could be a major bottleneck in the overall search if the individual search is performed by each CPU in parallel.
	These groups with low performance gains are not seen in the case combined with N-G, and thus {\em we suggest using all implementations in combination}.
	
\subsection[]{6D FoF}
\label{sec:Results-6DFoF}

\begin{figure*}
\centering
\includegraphics[width=0.95\textwidth]{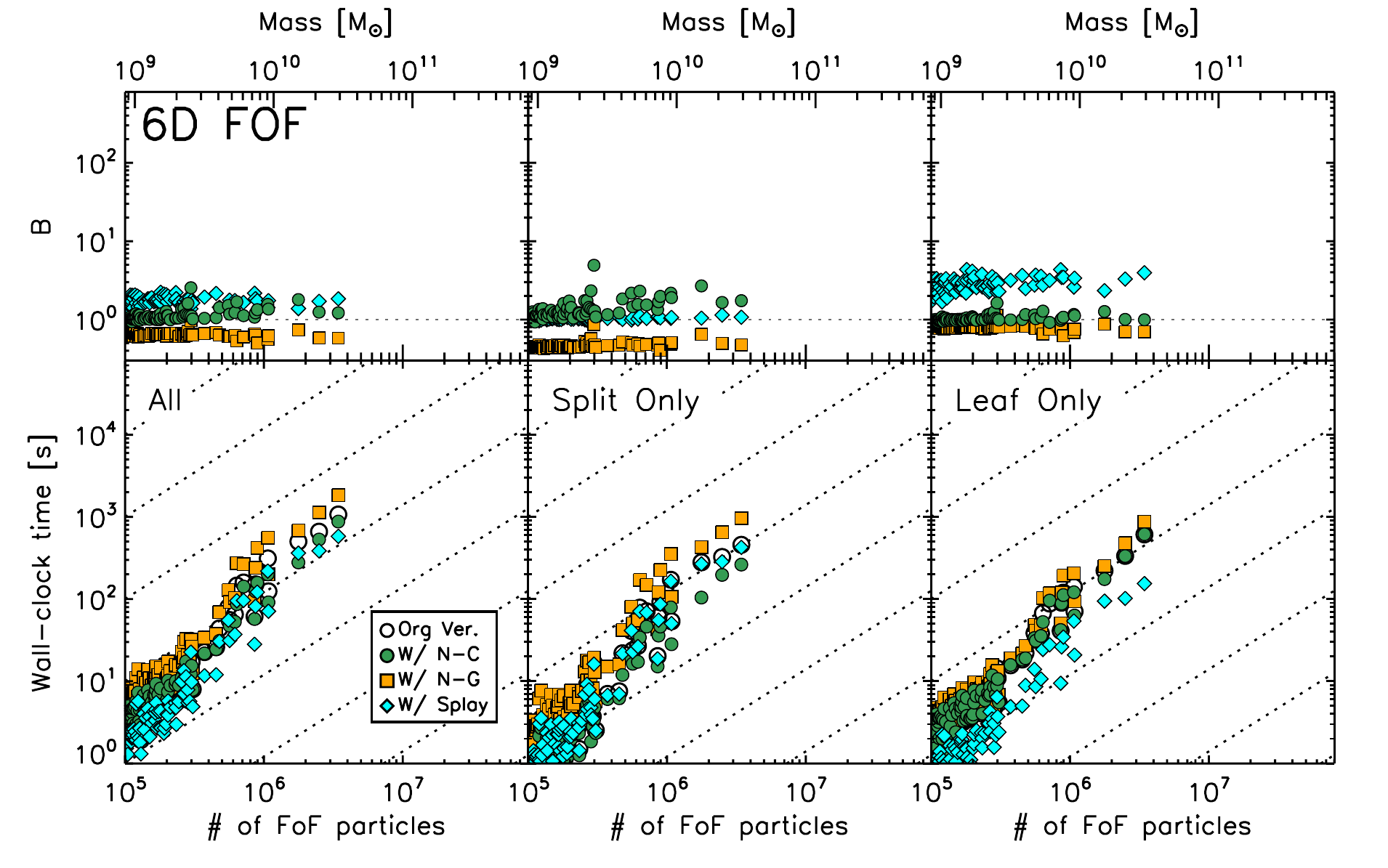}
\caption{6D FoF search results achieved with single implementations, with the same format as in Figure \ref{fig:fig Runtime_3D}.
Contrary to the 3D FoF results, the performance did not improve significantly.} 
\label{fig:fig Runtime_6D}
\end{figure*}

	We investigate the performance achievement of 6D FoF searches.
	Figure \ref{fig:fig Runtime_6D} illustrates the overall results of the 6D FoF searches in the same format as in Figure \ref{fig:fig Runtime_3D}.
	Compared with the 3D FoF search results, the 6D FoF search with each implementation shows smaller performance improvements.
	Moreover, the results with the N-G scheme show poorer wall-clock times than the original run, roughly by a factor of 1.5.
	Only the most massive 6D FoF group has a performance gain of $\sim 80 \%$ and $\sim 50\%$ in wall-clock times with N-C and Splay, respectively.
	Considering that the most massive FoF group is the major bottleneck in the overall execution time, such an improvement is desirable.
	However, the overall trend of wall-clock times departing from the ideal scaling for large 6D groups ($\gtrsim 10^{11}\,M_{\odot}$) demonstrates that the wall-clock time is too long for them.

	The improvement from N-C mainly occurs on the split nodes (bottom middle panel).
	This occurs mostly because the use of N-C significantly reduced the number of visits. 
	The improved efficiency of the 6D FoF search with N-C, however, is not as strong as that of the 3D FoF case.
	Most of the 6D FoF groups have a high fraction of their constituent nodes closed ($\sim 80\%$).
	This value is slightly smaller than that in the 3D FoF groups ($\sim 95\%$), but is still high.
	Surprisingly, the performance gain is not high on 6D searches with N-C: this means that a large number of visits are still required, even with a high fraction of closed nodes.
	We suspect that this is because the nodes are closed after multiple visits rather than a single visit in the case of high dimensions.
	This way, the N-C scheme has a lower efficiency for the FoF search in 6D, the so-called ``the curse of dimensionality''.
	This may be because nodes in 6D have a lower (phase-space) volume number density than nodes in 3D, and thus the rate of particle links per single visit gets lower in 6D.	
	
	To understand the curse of dimensionality, consider a cubic node with a side length of $2b$ (b is the linking length), in which particles are uniformly distributed.
	For a target particle located at the central position of the cube, the fraction of linked particles to all particles in the node can be approximated as the volume fraction of a sphere with radius $b$ to the volume of the cube: $f = \{\pi^{D/2}\,b^D / \Gamma\ (D/2 + 1) \} / (2b)^D$, $D$ being the dimension.
	This fraction converges to 0 as the dimension goes to infinity, implying that the efficiency of particle links gets lower at a higher dimension.
	For example, the fraction is 0.524 in 3D, which requires at least 2 visits for this node to be fully-connected, and 0.081 in 6D, corresponding to 13 visits at minimum.
	Therefore, in contrast with the 3D FoF search, a 6D node requires many more visits to be fully-connected.
	This explains why the N-C scheme shows low improvements in the 6D FoF search.

	Regarding N-G (orange squares in Figure \ref{fig:fig Runtime_6D}), the results show lower efficiencies---a factor of 1.5--- than those for the results without this scheme.
	Given that particles are more sparsely distributed in a higher dimension, the value of ${\rm max}\{ || \vec{X_{\rm i}} || \}$ in Equation \ref{eq:node geometry} gets bigger.
	Therefore, nodes in 6D are less frequently enclosed as the range of $|| \vec{D_{\rm i}} ||$ becomes wider.
	Without the benefits of the N-G scheme, additional calculations adversely affect the wall-clock times.

	The results with the Splay operation (cyan diamonds in Figure \ref{fig:fig Runtime_6D}) show the best efficiency of the three for the most massive group, but the performance is still poor.
	Given that hypercubic nodes are more difficult to be spanned with search spheres in higher dimensions, we suspect that the performance gains are related to the original purpose of the Splay operation combined with the default N-C scheme at the leaf node level: finding new particles efficiently.
	However, the overall scaling is still poor because the Splay operation alone does not improve the efficiency of neighbor queries.

\begin{figure}
\includegraphics[width=0.45\textwidth]{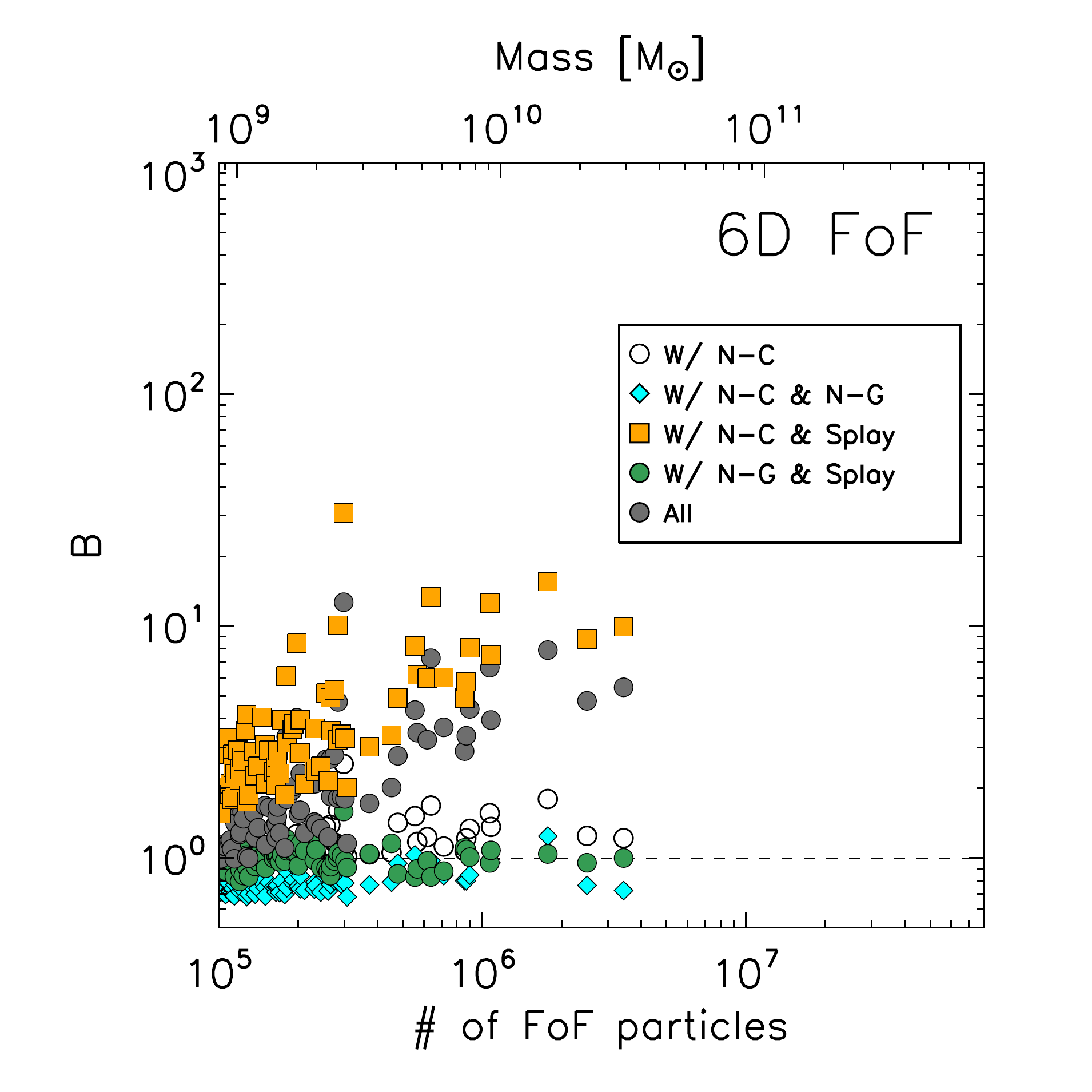}
\caption{
\BF\ parameter results for the 6D FoF groups in the same format as in Figure \ref{fig:fig Runtime_comb_3D}. 
Similar to the 3D FoF case, the combination of N-C and Splay has the best performance---more than 10 times faster than the original run.}
\label{fig:fig Runtime_comb_6D}
\end{figure}

	Again, we present the results for each combination of the three implementations in Figure \ref{fig:fig Runtime_comb_6D}.
	Of all combinations, applying both N-C and Splay operations (orange squares) shows the best results, with a speed being improved by a factor of 10 approximately ($\BF\ \sim 10$). 
	The performance gains are better for more massive FoF groups, making the overall wall-clock time close to the ideal value, as we aimed at with our initial implementations.
	The strong improvement achieved by the combination of the two implementations is attributed to the complementarity of the two implementations.
	N-C is effective for reducing the number of visits to already fully-connected nodes but not for finding new FoF particles, whereas Splay has the opposite functions.
	In other words, new particles of 6D FoF groups are effectively linked by Splay, making the constituent nodes fully-connected with fewer visits, and thus N-C becomes more efficient.
	
	With N-C and N-G combined (cyan diamonds), the efficiency of the 6D FoF search does not improve much, barely reaching $B \sim 1$.
	Similarly, the combination of N-G and Splay (green circles) shows a very minor improvement.
	It is probably the case that the benefits of Splay are lost in the additional calculations in N-G.
	Applying all implementations (gray circles) leads to significant improvement but is weaker than the best combination (N-C with Splay).
	However, it is still worth considering because N-G is effective if the FoF group has a particle bridge.
	With both N-C and Splay, the issue related to the curse of dimensionality seems to be solved at least up to 6D, and N-G may effectively find FoF groups with particle bridges.
	Therefore, in 6D FoF searches as well, {\em we propose using all implementations}, as in the 3D case.

\subsection[]{Tree Structure}
\label{sec:Results-Tree}

	In Section \ref{sec:Imple-Tree}, we introduced options that determine the overall structure of the KD tree.
	In this subsection, we aim to find the best options that yield the best FoF efficiency.
	The most basic premise for modifying the KD tree structure is to create a tree structure that maximally considers particle clustering as much as possible before performing an FoF search.
	In so doing, each node of the tree is likely to have particles with the same FoF group, and thus the likelihood of nodes being closed becomes higher compared with the original tree.

	As introduced in Section \ref{sec:Imple-Tree}, the KD tree has two properties: split dimension and split value.
	In this study, we use three methods to determine the split dimension: the dimension with the largest spread, the largest Shannon entropy, and the largest dispersion.
	In addition, we employ two methods to measure the split value: the median of particles' coordinates along the split dimension and the point at which the distance---projected to the split dimension--- to the closest particle becomes maximum.
	Admittedly, these are and only rely on naive ideas, and there could be more ingenious ways of building a tree.

\begin{figure}
\centering
\includegraphics[width=0.45\textwidth]{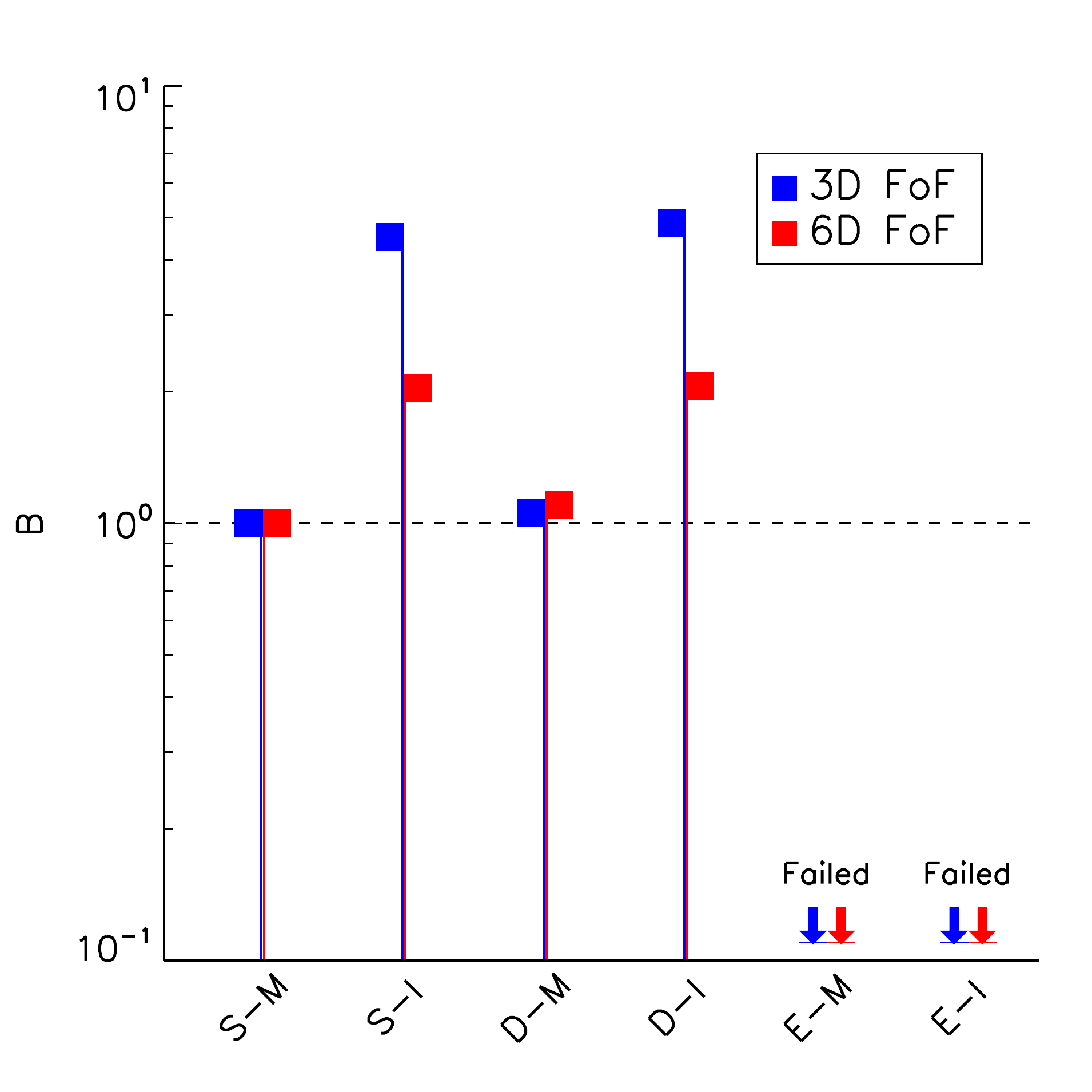}
\caption{
\BF\ parameter of the most massive FoF groups of each KD tree structure.
The original values for measuring the \BF\ parameters are obtained using all implementations and with the default KD tree structure.
Each KD tree structure is specified by a combination of split dimension and split value types.
The blue and red squares indicate the FoF search results for 3D FoF and 6D FoF, respectively.
The best FoF efficiency is achieved by the split value with the maximum inter-particle distance.}
\label{fig:fig Runtime_tree}
\end{figure}

	Figure \ref{fig:fig Runtime_tree} shows the results of the \BF\ parameter of the most massive 3D and 6D FoF groups for each KD tree structure.
	The results of the 3D FoF and 6D FoF searches are illustrated with blue and red squares, respectively.
	Each KD tree structure is labeled with a combination of the types of split dimension and split value.
	The split dimension with the largest spread, the largest dispersion, and the largest Shannon entropy is denoted ``S'', ``D'', and ``E'', respectively.
	In the same way, ``M'', and ``I'' denote the split value with the median of the coordinates and the maximum projected inter-particle distance, respectively.
	The \BF\ parameter has been measured in comparison with the default configuration that uses all three implementations (N-C, N-G, and Splay), the largest spread for the split dimension and the median for the split value (S-M), which is the set of values left-most in the figure.

	Among all KD tree structures, combinations containing the split values by the maximum inter-particle distance (S-I and D-I) show significant performance improvements.
	The best combination shows the search for the most massive 3D FoF group is 3.5 times faster than the reference run and 1.5 times faster for the most massive 6D FoF group. Much better results are obtained at a later epoch where particle clustering is more complicated---12.7 times faster for the 3D case and 2.7 times for the 6D case, as presented in Appendix \ref{sec:Appendix}.
	For example, with the D-I combination, the number of leaf nodes within the search sphere of the center particle of the most massive 3D FoF is 86,181 (c.f., 72,087 nodes in Panel (c) of Figure \ref{fig:fig GalandNode}).
	With the N-C scheme, only 5 open nodes are left (41 nodes remained with the default KD tree), which implies our purpose of {\textit pre-}considering clustering with the adaptive tree is well achieved.
	However, we note that it takes a slightly longer time to build an adaptive KD tree because a calculation is required to find the point with the maximum inter-particle distance\footnote{We simply sort particles along the split dimension and find the point with maximum inter-particle distance, which takes $O(N \log{N})$ computations.}.
	However, in all cases, the benefits of the adaptive KD tree overwhelm the extra time required to build the KD tree.

	No significant performance improvements were found through different choices of split dimension.
	With the largest Shannon entropy (E-M and E-I), the total wall-clock time increases by a factor of 100--1,000.
	Using the largest dispersion (D-M and D-I) does not provide any speed boost compared with the reference run.
	Thus far, the best combination for constructing a KD tree is to use split dimensions with the largest spread or the maximum dispersion and split values with the maximum inter-particle distance (S-I or D-I).
	Calculations to find these values are inexpensive, and the benefit of using them in FoF searches is clearly significant. 
	
	Furthermore, the adaptive KD tree can be applied to decompose the simulation into domains.
	Many FoF galaxy finder codes decompose the entire simulation volume for parallelization.
	However, a naive volume-based parallelization causes a large FoF group (e.g., a galaxy) to be redundantly identified as local FoF groups in multiple neighboring domains, and merging these FoF groups into one FoF is computationally very expensive.
	
	An adaptive KD tree with additional conditions can be used for economic domain decomposition, to avoid this problem.
	When building an adaptive KD tree from the entire domain, we add a condition in which the inter-particle distance for determining split values must be larger than the linking length.
	The sizes of leaf nodes depend on the purpose of the study\footnote{A typical galaxy study may set the maximum size of a leaf node to be $10^{7}$ particles, corresponding to $10^{11} M_{\odot}$.
	Conversely, we have a leaf node containing at least a certain number of particles (1M) to avoid creating an overly large number of domains.}.
	Then, each leaf node can be considered a decomposed domain.
	This implementation decomposes the entire domain into multiple subdomains, avoiding the large computations necessary for merging local FoF groups and providing a high level of parallel scalability.
	However, we note that building the new KD tree introduces the increase in computational cost.
	For example, KD tree with the new split value definition (I type) requires computing the distances of all particles to their closest one, and thus the wall-clock time for tree building gets longer.
	But, we note that the benefits from the new tree structure to the efficiency of FoF search and the domain-stitching overwhelm the extra building time for tree itself in most of the cases.

\section[]{Conclusion}
\label{sec:Conclusion}

The FoF method has been widely used to find galaxies and halos in numerical simulations, mostly because of its simplicity and efficiency. 
However, the standard FoF algorithm can become unacceptably slow for very high resolution hydrodynamical simulations, which have star particles that are significantly more clustered than the dark matter particles.
We confirm this using the 6D FoF galaxy finder, \VR, on the recent high-resolution simulation, \NH. 
The main reason for the poor performance of the FoF search is the large number of computations in the central regions of galaxies (Section \ref{sec:Background}).
High number density peaks result in the tree structure being too deep, resulting in a large number of visits to the tree nodes (Figure \ref{fig:fig Vtimes_All}).
Hence, the number density in the central region (i.e., particle resolution) is a key parameter that determines the FoF performance.

To overcome this problem, we develop simple modifications that can be readily applied to any FoF galaxy finder using the KD tree scheme.
First, we propose two methods of lowering the depth of the tree structure during FoF searches.
One is to close a fully-connected node such that the node is no longer visited (N-C; Section \ref{sec:Imple-NC}), and the other is to skip visiting a node based on the geometry between the node and the target particle (N-G; Section \ref{sec:Imple-NG}).
Furthermore, we present the Splay operation (Section \ref{sec:Imple-Splay}), which more efficiently finds the member particles of an FoF group, as well as the shape of the KD tree structure with a high FoF search efficiency (Section \ref{sec:Imple-Tree}).
All proposed schemes show a remarkable performance improvement with no or little extra memory usage, as summarized below.

\begin{figure}
\includegraphics[width=0.45\textwidth]{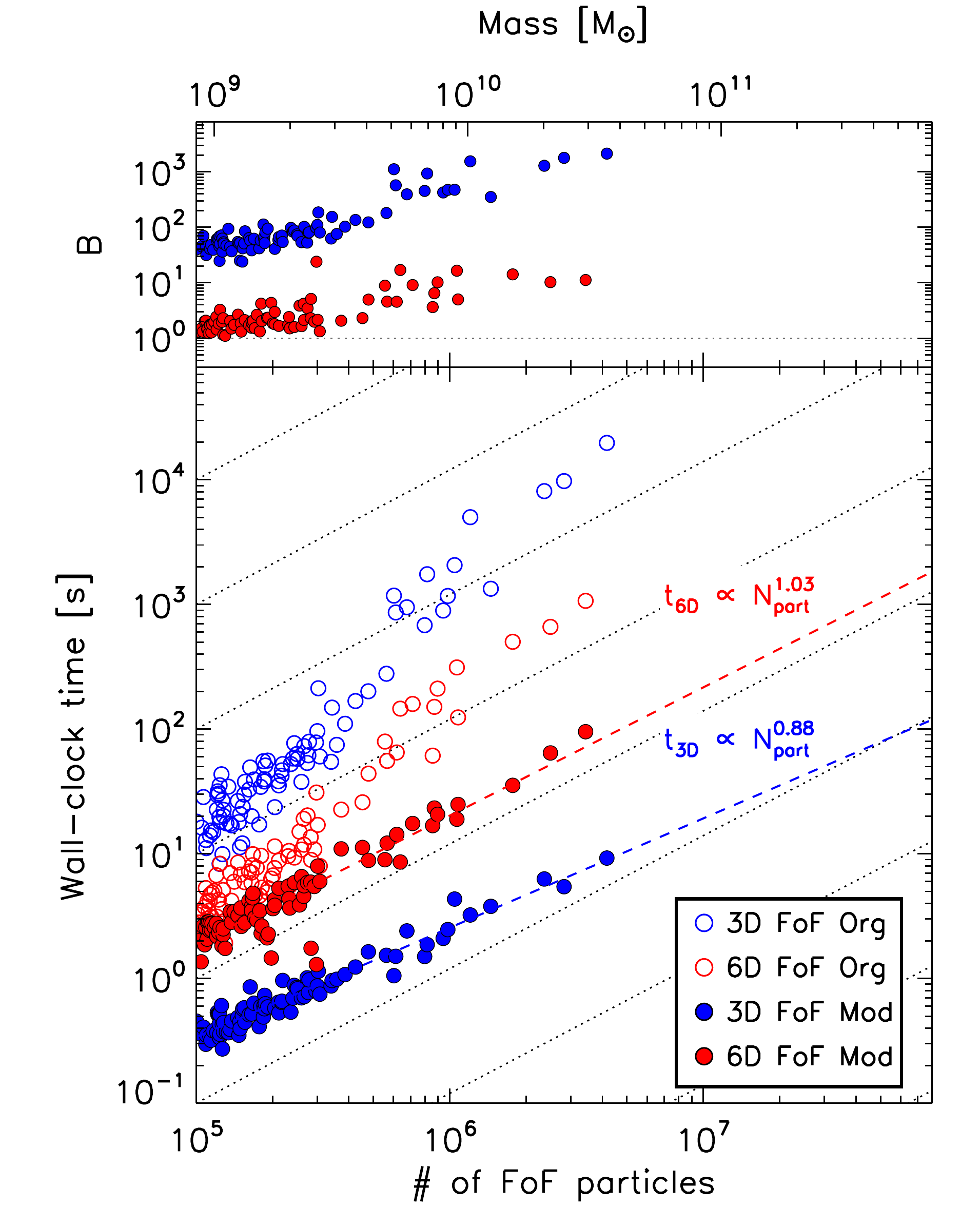}
\caption{Achieved FoF performance results of applying all implementations.
In the bottom panel, the open circles are the results of the original run (the same as in Figure \ref{fig:fig Runtime_All}), and the closed circles are the results achieved.
Red circles correspond to the results of the 3D FoF searches and blue circles for the 6D case.
The results show significant performance improvements, in which the scaling is very comparable with the ideal scaling (dotted lines).
In the upper panel, the \BF\ parameter results with respect to the original ones are plotted.
For the most massive 3D FoF group (or 6D FoF group), the search speed is improved by a factor of approximately 2700 (or 12).
}
\label{fig:fig Runtime_best}
\end{figure}

\begin{table*}
  \centering
  \caption{Overall run report.}
  \begin{tabular}{cccccc}
  \hline \hline
Runtype  & \multicolumn{3}{c}{Wall-clock time in each routine [sec]} &  \multicolumn{2}{c}{Memory usage [GB]}  \\ [5pt]
  & Read data \tablenotemark{\it a} & 3D FoF \tablenotemark{\it b} & 6D FoF  \tablenotemark{\it c} & Average \tablenotemark{\it d} & Peak \tablenotemark{\it e} \\ [5pt]
  \hline
 Original ver. & 221 & ${19,767}$ & ${1,069}$ & 8.76 & 12.01 \\ [3.5pt]
 With all implementations & 221 & 9 & 95 & 9.36 & 12.70 \\ [3.5pt]
 \hline

 \end{tabular}
 \label{tab:report}
 
\raggedright
\tablenotetext{a}{Reading the data with $47,218,669$ particles.}
\tablenotetext{b}{Wall-clock time for searching the most massive 3D FoF object.}
\tablenotetext{c}{Wall-clock time for searching the most massive 6D FoF object.}
\tablenotetext{d}{Average memory usage during the execution.}
\tablenotetext{e}{Peak memory usage during the execution.}
 
\end{table*}

\begin{itemize}
	\item
	Regarding 3D FoF searches (Section \ref{sec:Results-3DFoF}), a single implementation of N-C or N-G shows significant performance improvements (Figure \ref{fig:fig Runtime_3D}).
	This implies that the slowdown of the 3D FoF searches is due to the repetitive visits to already fully-connected nodes.
	Of all sets of associated implementations, the combination of N-C and Splay shows the best performance for most massive FoF groups (Figure \ref{fig:fig Runtime_comb_3D}).
	However, a few FoF groups in which structures are connected with a particle bridge do not have any performance gain with this combination.
    Only when the N-G scheme is applied together with both N-C and Splay do all 3D FoF groups show a performance improvement.
	Thus, we propose applying all schemes for 3D FoF searches.
	
	\item
	In contrast, our 6D FoF searches did not show a notable performance when we used any one of the three implementations alone (Figure \ref{fig:fig Runtime_6D} in Section \ref{sec:Results-6DFoF}).
	While some massive 6D FoF groups show mildly reduced wall-clock times with N-C or Splay, the overall scaling implies that more massive galaxies at later epochs will take too much time to search.
	We suspect that the main reason for the poor performance is that the current KD tree scheme does not show optimal performance in a high dimension, commonly referred to as the curse of dimensionality.
	In other words, the fraction of a search sphere to the volume of a single node is significantly reduced at high dimensions, and thus the efficiency of a single neighbor query is significantly reduced.
	
	\item
	A significant improvement in the 6D FoF searches is achieved when both N-C and Splay are simultaneously implemented (Figure \ref{fig:fig Runtime_comb_6D}).
	These schemes work complementarily, overcoming the curse of dimensionality problem at least in part.
	We again suggest using all three schemes to avoid poor performance on some FoF groups with particle bridges.
	
	\item
	We tested different KD tree structures by adopting different ways of defining split dimensions and split values that allow non-balanced trees as well (Section \ref{sec:Results-Tree}).
	The most significant improvement in performance is achieved using a split value at which the distance to the closest particle (in the projected space to the split dimension) is maximized.
	The choice of split dimension with the largest dispersion also shows mild improvement.
	FoF searches with modified KD tree structures achieve a speed boost of a factor of 3.5 (or 1.5) in 3D (or 6D) FoF searches (Figure \ref{fig:fig Runtime_tree}).
	Furthermore, the proposed KD tree structure can be applied to decompose domains, avoiding additional expensive calculations when stitching an FoF group across the boundary of domains.
	
	\item
	Using all implementations brings a significant speed-up of the overall execution: $>1,000$ (or $10$) times faster than before for the 3D (or 6D) FoF search (see Figure \ref{fig:fig Runtime_best}).
	The degree of the boost is much more remarkable for a snapshot with more massive galaxies (see Figure \ref{fig:fig Runtime_best_400}).
	Both the improved 3D and 6D FoF searches reach the ideal scaling as desired.
	However, even with the updated implementations, the 6D FoF search is approximately 10 times slower than the 3D FoF search.
	Thus, the 6D FoF search is particularly difficult because the extra dimensions of velocity are clustered in a completely different fashion to configuration-space.
	
	\item
	Table \ref{tab:report} shows the overall run report.
	We observe that there is no or little extra usage of other computing resources (e.g., memory) with our implementations.
	Therefore, the reduced wall-clock time effectively indicates the performance enhancement without trade-off with other computing resources.
	
\end{itemize}

{\em The new implementations resulted in the identical halo and galaxy detection to the original run}. Hence, scientific analysis is not affected by the new implementations.
Figure \ref{fig:fig Runtime_best} shows the final results of the FoF search performance.
The open circles represent the original runs (the same as those in Figure \ref{fig:fig Runtime_All}), and the closed circles are the results of applying all implementations and with the best KD tree structure.
The final results are very comparable to the ideal scaling relations (dotted lines).
Utilizing these scaling relations, an FoF search for a massive galaxy can be completed within a day, even when the particle resolution in the simulation becomes $100 M_{\odot}$.
We also emphasize that our schemes find a {\em unique} solution of FoF groups within a practicable wall-clock time.
Hence, the schemes can be used for various types of identification, such as using different types of particles with different linking lengths.

Finally, we have applied the updated version of \VR\footref{VRgithub} on the entire \NH\ simulation (834 consecutive snapshots: $a = 0.091 - 0.853$) and identified its galaxies and halos.
The merger tree of halos and galaxies have been built with {\textsc{TreeFrog}} \citep[][]{Elahi19b}.
An open IDL library is used for calculating integrated properties of halos/galaxies with \VR\ output and post-processing the \textsc{TreeFrog} output, in which some useful calling routines are included\footnote{\dataset[https://github.com/JinsuRhee/VELOCIraptor{\_}IDL{\_}Tools]{https://github.com/JinsuRhee/VELOCIraptor_IDL_Tools}}.
A series of scientific reports using these output will follow this paper.
Furthermore, considering the wide range of galaxy properties (e.g., mass), there is no universal choice of linking length for identifying substructures.
In another technical paper, we will tackle this issue in detail.

\acknowledgments
S.K.Y. acknowledges support from the Korean National Research Foundation (NRF-2020R1A2C3003769).
The supercomputing time for numerical simulation was kindly provided by KISTI (KSC-2017-G2-003, KSC-2019-CHA-0006), and large data transfer was supported by KREONET, which is managed and operated by KISTI.
As the head of the group, S.K.Y. acted as the corresponding author.
We thank Juhan Kim for his valuable comments during the early phases of the investigation.


\appendix

\renewcommand\thefigure{\thesection\arabic{figure}} 
\setcounter{figure}{0}
\renewcommand\thetable{\thesection\arabic{table}} 
\setcounter{table}{0}

\section{Run report with a different snapshot}
\label{sec:Appendix}

\begin{table*}
  \centering
  \caption{Overall run report for another snapshot.}
  \begin{tabular}{cccccc}
  \hline \hline
Runtype  & \multicolumn{3}{c}{Wall-clock time in each routine [sec]} &  \multicolumn{2}{c}{Memory usage [GB]}  \\ [5pt]
  & Read data \tablenotemark{\it a} & 3D FoF \tablenotemark{\it b} & 6D FoF  \tablenotemark{\it c} & Average \tablenotemark{\it d} & Peak \tablenotemark{\it e}\\ [5pt]
  \hline
 Original ver. & 748 & ${1,043,249}$ & ${83,044}$ & 24.84 & 53.09 \\ [3.5pt]
 With all implementations & 706 & 39 & 355 & 24.55 & - \\ [3.5pt]
 \hline

 \end{tabular}
 \label{tab:report400}
 
\raggedright
\tablenotetext{a}{Reading the data with $174,102,517$ particles.}
\tablenotetext{b}{Wall-clock time for searching the most massive 3D FoF object.}
\tablenotetext{c}{Wall-clock time for searching the most massive 6D FoF object.}
\tablenotetext{d}{Average memory usage during the execution.}
\tablenotetext{e}{Peak memory usage during the execution.}
 
\end{table*}

\begin{figure*}
\centering
\includegraphics[width=0.95\textwidth]{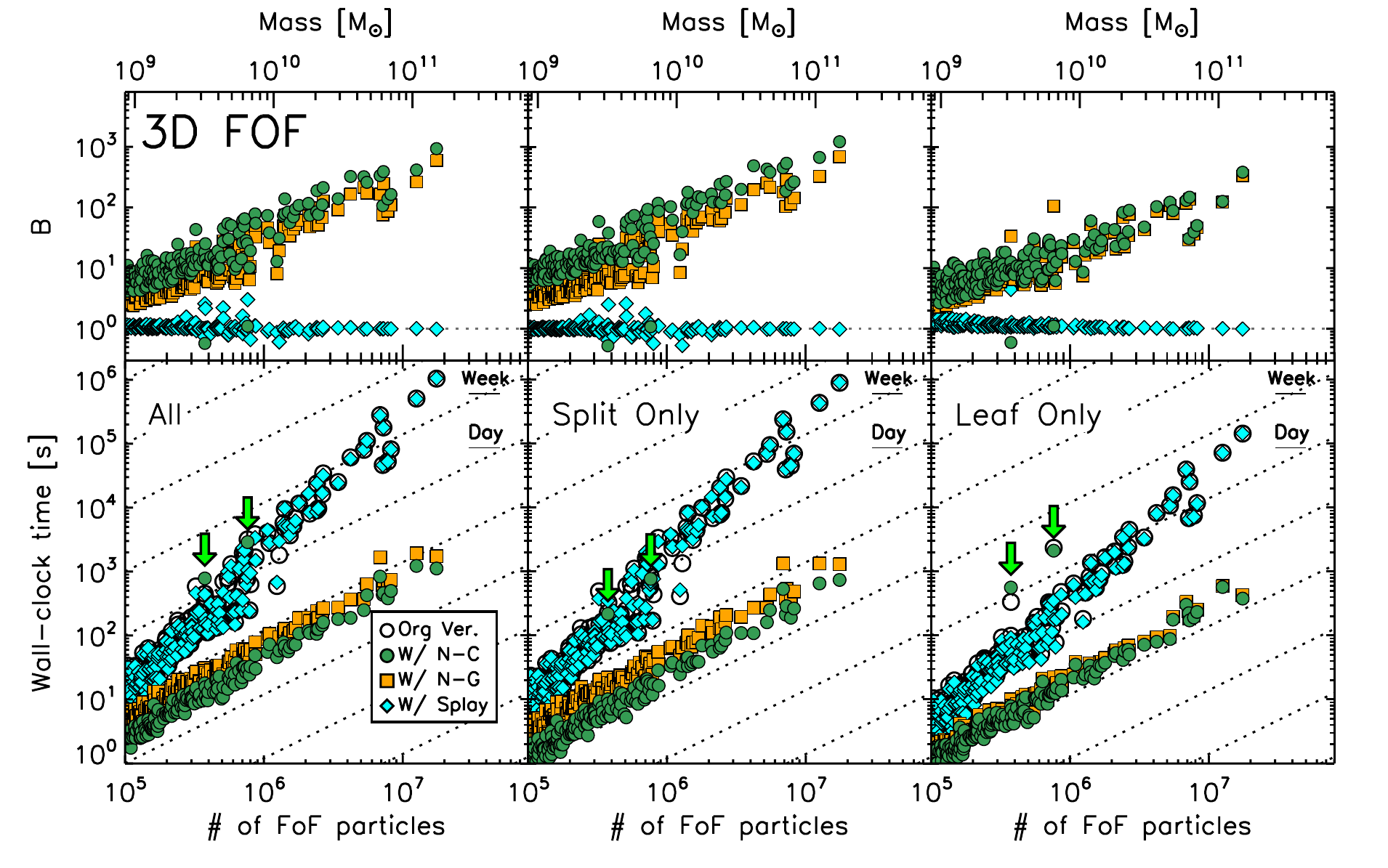}
\caption{Same format as in Figure \ref{fig:fig Runtime_3D} but with the snapshot data at $a = 0.462$.}
\label{fig:fig Runtime_3D_400}
\end{figure*}

\begin{figure*}
\centering
\includegraphics[width=0.95\textwidth]{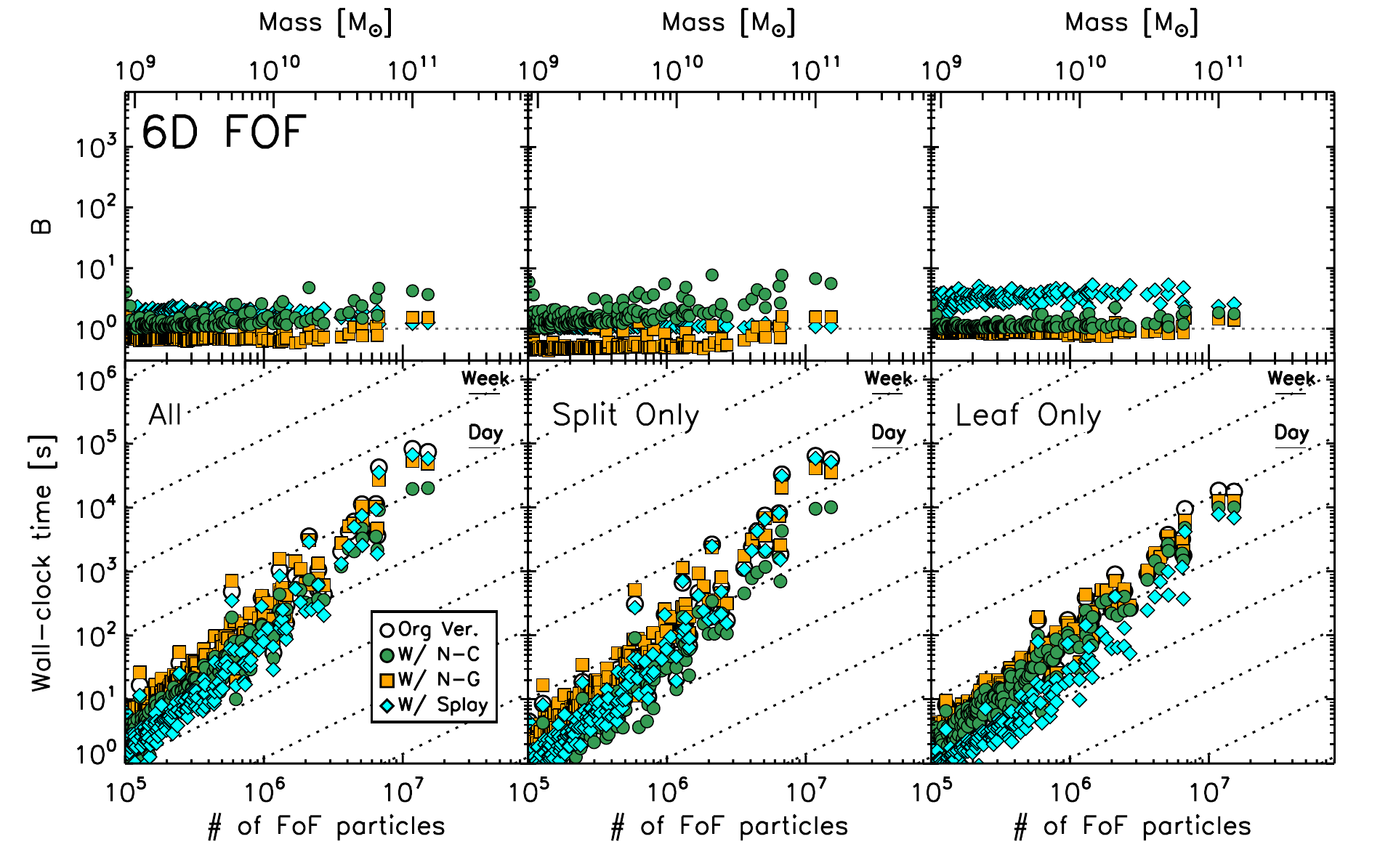}
\caption{Same format as in Figure \ref{fig:fig Runtime_6D} but with the snapshot data at $a = 0.462$.}
\label{fig:fig Runtime_6D_400}
\end{figure*}

\begin{figure*}
\centering
\includegraphics[width=0.75\textwidth]{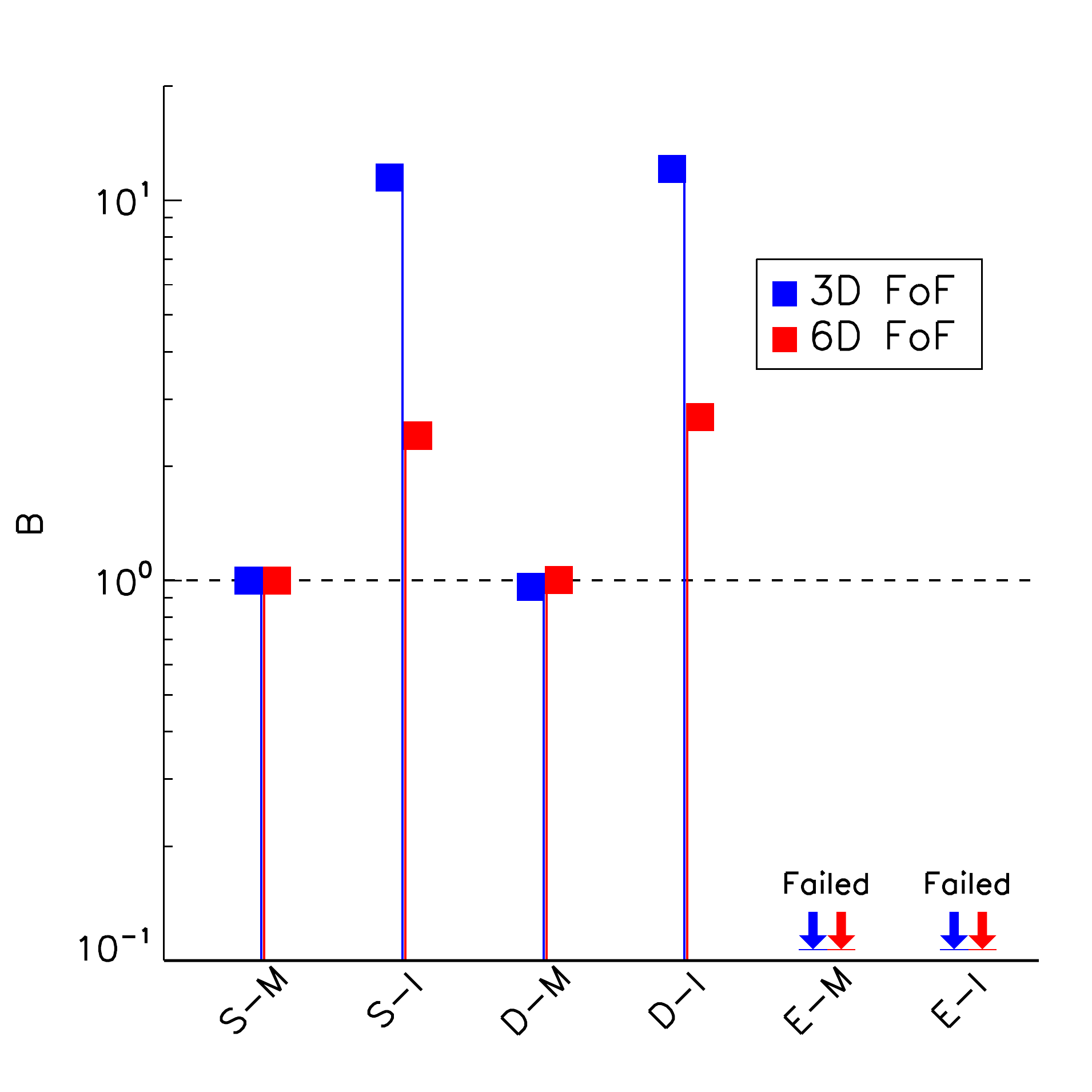}
\caption{Same format as Figure \ref{fig:fig Runtime_tree} but with the snapshot data at $a = 0.462$.}
\label{fig:fig Runtime_tree_400}
\end{figure*}

\begin{figure*}
\centering
\includegraphics[width=0.75\textwidth]{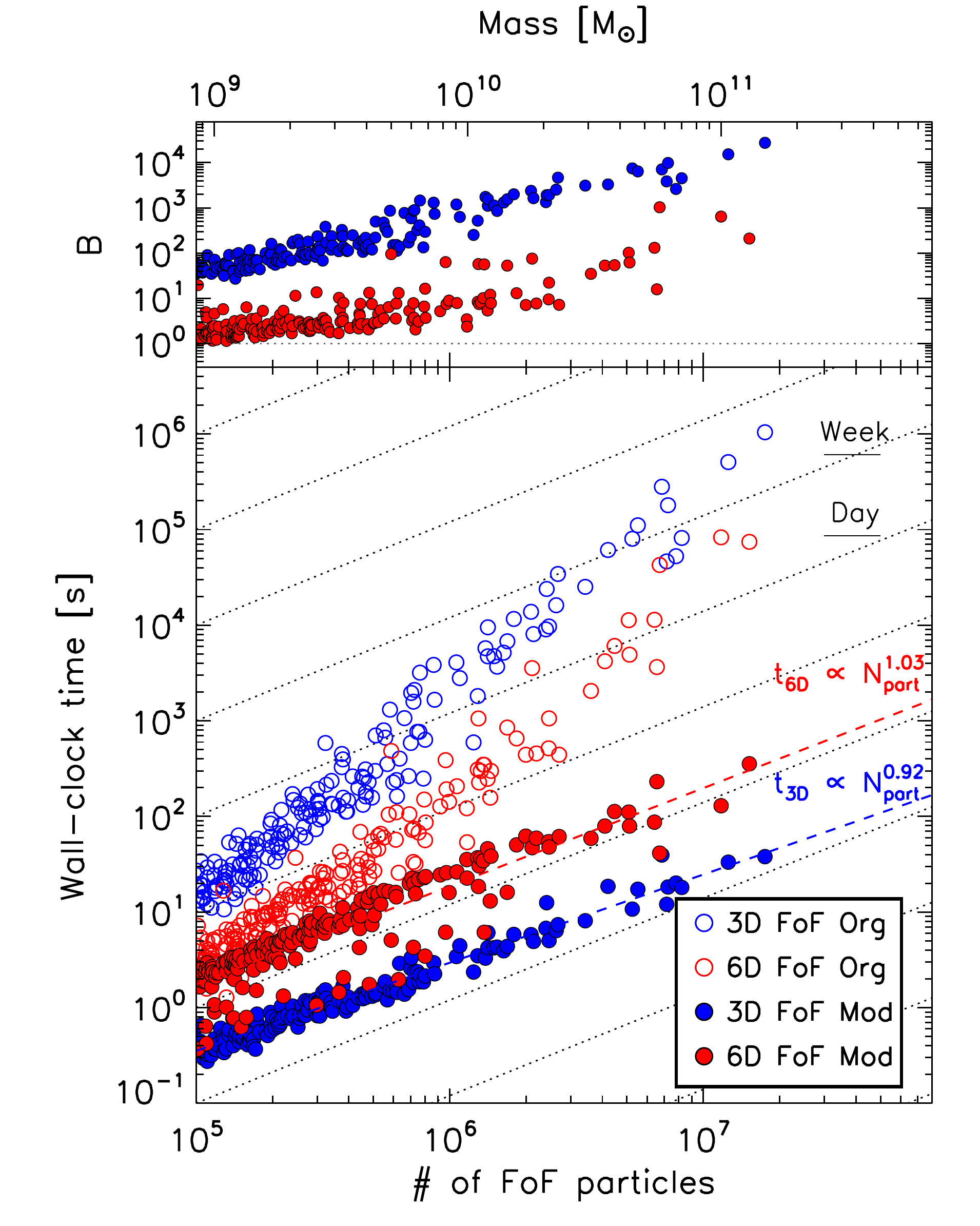}
\caption{Same format as in Figure \ref{fig:fig Runtime_best} but with the snapshot data at $a = 0.462$.}
\label{fig:fig Runtime_best_400}
\end{figure*}

As mentioned in the main text, we applied the new code to the entire snapshots of the \NH\ simulation.
In this section, we report on the performance improvement observed in the snapshot of the \NH\ simulation ($a = 0.462$), at which typical massive galaxies ($\gtrsim 10^{11}\,M_{\rm \odot}$) are being formed.
The run report of this snapshot is given in Table \ref{tab:report400} in the same format as Table \ref{tab:report}. The peak memory usage with the original version cannot be measured due to its impractically-long execution time. Again, we do not see any extra memory usage or I/O time with our new implementations.
Figure \ref{fig:fig Runtime_3D_400}, \ref{fig:fig Runtime_6D_400}, \ref{fig:fig Runtime_tree_400} and \ref{fig:fig Runtime_best_400} have the exact same format as in Figure \ref{fig:fig Runtime_3D}, \ref{fig:fig Runtime_6D}, \ref{fig:fig Runtime_tree}, and \ref{fig:fig Runtime_best}, respectively, but for the different snapshot.
As the poor scaling seen with the original snapshot data ($a = 0.289$) suggested, finding the most massive 3D (or 6D) groups takes more than a week (or a day for 6D), respectively (see Figure \ref{fig:fig Runtime_3D_400} and \ref{fig:fig Runtime_6D_400}).
In particular, as demonstrated in Section \ref{sec:Results-3DFoF}, 3D FoF groups without the N-G scheme (arrowed green circles in Figure \ref{fig:fig Runtime_3D_400}) show no performance improvement.
The wall-clock time for the more massive one of the two exceeds that of the most massive 3D FoF group, which is the bottleneck in the overall execution in a parallel manner.
Figure \ref{fig:fig Runtime_best_400} repeatedly highlights the success of this study, achieving the ideal scaling relations.

\begin{thebibliography}{}
\bibitem[Aubert et al.(2004)]{Aubert04} Aubert, D., Pichon, C., \& Colombi, S. 2004, \mnras, 352, 376
\bibitem[Bakels et al.(2021)]{Bakels21} Bakels, L., Ludlow, A. D., Power, C., et al. 2021, \mnras, 501, 5948
\bibitem[Bentley (1975)]{Bentley75} Bentley, J. L., 1975, Common. ACM,18, 509
\bibitem[Behroozi et al.(2013)]{Behroozi13} Behroozi, P. S., Wechsler, R. H., \& Wu, H. 2013, \apj, 762, 109
\bibitem[Borrow et al.(2020)]{Borrow20} Borrow, J., Angl\'{e}s-Alc\'{a}zar, D., \& Dav\'{e}, R. 2020, \mnras, 491, 6102
\bibitem[Ca\~{n}as et al.(2019)]{Canas19} Ca\~{n}as, R., Elahi, P. J., Welker, C., et al. 2019, \mnras, 482, 2039
\bibitem[Cormen et al.(2009)]{Cormen09} Cormen, T. H., Leiserson, C. E., Rivest, R. L., et al. 2009, Introduction to Algorithms Third edition, The MIT press
\bibitem[Creasey (2018)]{Creasey18} Creasey, P. 2018, A\&C, 25, 159
\bibitem[Davis et al.(1985)]{Davis85} Davis, M., Efstathiou, G., Frenk, C. S., \& White, S. D. M. 1985, \apj, 292, 391
\bibitem[Dolag et al.(2009)]{Dolag09} Dolag, K., Borgani, S., Murante, G., \& Springel, V., 2009, \mnras, 399, 497
\bibitem[Dubois et al.(2014)]{Dubois14} Dubois, Y., Pichon, C.; Welker, C., et al. 2014, \mnras, 444, 1453
\bibitem[Dubois et al.(2021)]{Dubois21} Dubois, Y., Beckmann, R., Bournaud, F., et al. 2021, \aap, 651, 109
\bibitem[Elahi et al.(2018)]{Elahi18} Elahi, P. J., Welker, C., Power, C., et al. 2018, \mnras, 475, 5338
\bibitem[Elahi et al.(2019a)]{Elahi19a} Elahi, P. J., Ca\~{n}as, R., Poulton, R. J. J., et al. 2019a, \pasa, 36, 21
\bibitem[Elahi et al.(2019b)]{Elahi19b} Elahi, P. J., Poulton, R. J. J., Tobar, R. J., et al. 2019b, \pasa, 36, 28
\bibitem[Feng \& Modi (2017)]{FM17} Feng, Y., \& Modi, C. 2017, Astronomy and Computing, 20, 44
\bibitem[Gelb \& Bertschinger (1994)]{GB94} Gelb, J. M., Bertschinger, E., 1994 \apj, 436, 467
\bibitem[Hopkins et al.(2018)]{Hopkins18} Hopkins, P. F., Wetzel, A., Kere\v{s}, D., et al. 2018, 480, 800
\bibitem[Knebe et al.(2011)]{Knebe11} Knebe, A., Knollmann, S. R., Muldrew, S. I., et al. 2011, \mnras, 415, 2293
\bibitem[Knebe et al.(2013a)]{Knebe13a} Knebe, A., Libeskind, N. I., Pearce, F., et al. 2013a, \mnras, 428, 2039
\bibitem[Knebe et al.(2013b)]{Knebe13b} Knebe, A., Pearce, F. R., Lux, H., et al. 2013b, \mnras, 435, 1618
\bibitem[Knollmann \& Knebe (2009)]{KK09} Knollmann S. R., \& Knebe, A., 2009, \apjs, 182, 608
\bibitem[Komatsu et al.(2011)]{Komatsu11} Komatsu, E., Smith, K. M., Dunkley, J., et al. 2011, \apjs, 192, 18
\bibitem[Kwon et al. (2010)]{Kwon10} Kwon, Y., Nunley., D., Gardner, J. P., et al. 2010, Scientific and Statistical Database Management pp 132-150, Springer Berlin Heidelberg
\bibitem[Moore et al.(2001)]{Moore01} Moore, A. W. Connolly, A. J., Genovese, C., et al. 2001, Mining the Sky: Proceedings of the MPA/ESO/MPE Workshop, p. 71
\bibitem[More et al.(2011)]{More11} More, S., Kravtsov, A. V., Dalal, N., \& Gottl\"{o}ber, S., 2011, \apjs, 195, 4
\bibitem[Power et al.(2020)]{Power20} Power, C., Elahi, P. J., Welker, C., et al. 2020, \mnras, 491, 3923
\bibitem[Press \& Schechter (1974)]{PS74} Press, W. H., \& Schechter, P. 1974, \apj, 187, 425
\bibitem[Roy et al.(2014)]{Roy14} Roy, F., Bouillot, V. R., \& Rasera, Y. 2014, \aap, 564, 13
\bibitem[Schaye et al.(2015)]{Schaye15} Schaye, J., Crain, R. A., Bower, R. G., et al. 2015, \mnras, 446, 521
\bibitem[Springel et al.(2001)]{Springel01} Springel, V., White, S. D. M., Tormen, G., \& Kauffmann, G. 2001, \mnras, 328, 726
\bibitem[Springel et al.(2005)]{Springel05a} Springel, V., White, S. D. M., Jenkins, A., et al. 2005, \nat, 435, 629
\bibitem[Springel (2005)]{Springel05b} Springel, V. 2005, \mnras, 364, 1105
\bibitem[Springel et al.(2021)]{Springel21} Springel, V., Pakmor, R., Zier, O., \& Reinecke, M. 2021, \mnras, 506, 2871
\bibitem[Stadel (2001)]{Stadel01} Stadel, J. G., 2001, PhD thesis, AA(UW)
\bibitem[Teyssier (2002)]{Teyssier02} Teyssier, R. 2002, \aap, 385, 337
\bibitem[Vogelsberger et al.(2014)]{Vogelsberger14} Vogelsberger, M., Genel, S., Springel, V., et al. 2014, \mnras, 444, 1518
\bibitem[Wright et al.(2020)]{Wright20} Wright, R. J., Lagos, C. d. P., Power, C., et al. 2020, \mnras, 498, 1668
\end{thebibliography}
\end{document}